\documentclass[12pt,journal,onecolumn,draftclsnofoot]{IEEEtran}
\usepackage[left=1.9cm, right=1.9cm, top=1.9cm, bottom = 1.9cm]{geometry}
\usepackage[T1]{fontenc}
\usepackage[utf8]{inputenc}
\usepackage{authblk}
\usepackage{amsthm}
\usepackage{cite}
\include{psfig}
\usepackage{psfig}
\usepackage{setspace}
\usepackage{psfrag}
\usepackage{amssymb}
\usepackage{picins}
\usepackage[demo]{graphicx}
\graphicspath{ {images/} }
\usepackage{wrapfig}
\usepackage{graphicx}
\usepackage{tikz}
\usepackage{multicol,lipsum}
\usepackage{subcaption}
\usepackage{enumitem}
\usepackage{tikz}
\usetikzlibrary{calc}
\usepackage[tbtags]{amsmath}
\usepackage{epsfig}
\usepackage{epstopdf}
\usepackage{stfloats}
\usepackage{float}
\usepackage{array}
\usepackage{color}
\usepackage{pgfplots}
\usepackage{textcomp}
\usepackage{gensymb}
\usepackage{multirow}
\usepackage{url}
\usepackage{comment}

\usepackage[ruled,vlined]{algorithm2e}

\usepackage{algorithmic}
\usepackage{adjustbox}
\usepackage{lipsum}
\pgfplotsset{compat=newest}
\pgfplotsset{plot coordinates/math parser=false}
\usepackage{stmaryrd}

\IEEEoverridecommandlockouts
\newlength\tindent
\renewcommand{\indent}{\hspace*{\tindent}}

\def\BibTeX{{\rm B\kern-.05em{\sc i\kern-.025em b}\kern-.08em
  T\kern-.1667em\lower.7ex\hbox{E}\kern-.125emX}}

\begin{document}
\title{\huge{Joint Resource Allocation and Phase Shift Optimization for RIS-Aided eMBB/URLLC Traffic Multiplexing}}
\author{Mohammed AL-Mekhlafi, Mohamed Amine Arfaoui, Mohamed Elhattab, Chadi Assi, and Ali Ghrayeb }
\maketitle
\begin{abstract}
This paper studies the coexistence of enhanced mobile broadband (eMBB) and ultra-reliable and low-latency communication (URLLC) services in a cellular network that is assisted by a reconfigurable intelligent surface (RIS). The system model consists of one base station (BS) and one RIS that is deployed to enhance the performance of both eMBB and URLLC in terms of the achievable data rate and reliability, respectively. We formulate two optimization problems, a time slot basis eMBB allocation problem and a mini-time slot basis URLLC allocation problem. The eMBB allocation problem aims at maximizing the eMBB sum rate by jointly optimizing the power allocation at the BS and the RIS phase-shift matrix while satisfying the eMBB rate constraint. 
On the other hand, the URLLC allocation problem is formulated as a multi-objective problem with the goal of maximizing the URLLC admitted packets and minimizing the eMBB rate loss. This is achieved by jointly optimizing the power and frequency allocations along with the RIS phase-shift matrix. In order to avoid the violation in the URLLC latency requirements, we propose a novel framework in which the RIS phase-shift matrix that enhances the URLLC reliability is proactively designed at the beginning of the time slot. For the sake of solving the URLLC allocation problem, two algorithms are proposed, namely, an optimization-based URLLC allocation algorithm and a heuristic algorithm. The simulation results show that the heuristic algorithm has a low time complexity, which makes it practical for real-time and efficient multiplexing between eMBB and URLLC traffic. In addition, using only 60 RIS elements, we observe that the proposed scheme achieves around 99.99\% URLLC packets admission rate compared to 95.6\% when there is no RIS, while also achieving up to 70\% enhancement on the eMBB sum rate.
\end{abstract}
\section{Introduction}
\subsection{Motivation and Background}
\textcolor{black}{\indent The upcoming sixth generation (6G) wireless networks are expected to support a wide range of new applications and services beyond those supported by the current fifth generation (5G) wireless networks, namely, enhanced mobile broadband (eMBB), ultra-reliable and low latency communications (URLLC) and massive machine type communications (mMTC) \cite{8869705,9170653,8808168}. These emerging applications demand enhanced and stricter requirements than those supported by 5G \cite{park2020extreme}. Additionally, multiple requirements may be concurrently demanded by some applications. For instance, high reliability, low latency and high data rates are services which all are needed for enabling extended reality (XR), which is one of the visioned applications of 6G networks  \cite{park2020extreme,8808168}. Consequently, the coexistence of different service classes, especially the eMBB and URLLC services, within the same time/frequency resources, is a major challenge that already exists in 5G networks and that will escalate in the upcoming 6G networks \cite{park2020extreme,bennis_risk,9279247}. This is mainly due to the different requirements of each service, for instance eMBB service focuses on transmitting large blocks of data with high data rates, while the URLLC service targets short packets transmission with an extremely high reliability and a low end-to-end latency \cite{park2020extreme,bennis_risk}.\\
\indent Resource puncturing has been recently proposed as an enabler for the coexistence of eMBB and URLLC services in the same network \cite{pedersen2017punctured,joint}. This method consists of puncturing (e.g., allocating no power on) some frequency resources which have already been allocated to the eMBB load and assigning such resources to the URLLC service \cite{joint}. While this enables the transmission of the URLLC traffic along with the eMBB load, it may have adverse impacts on the performance and the spectral efficiency of the eMBB service, since loosing spectrum resources to favor the URLLC service will reduce the data and transmission rates of the eMBB traffic \cite{joint,8403963_Physical}. Now, the majority of eMBB applications have target quality-of-service (QoS) to satisfy, e.g., minimum data rates. Such constraint directly relates to the required number of frequency resources which should be allocated to each eMBB user. Accordingly, the maximum URLLC load that can be transmitted simultaneously with the ongoing eMBB load using the puncturing scheme is limited by the QoS constraint of each eMBB user.  \\
\indent The performance of the eMBB and URLLC services, which is measured in terms of data rates for the eMBB traffic and the reliability and latency for the URLLC traffic, depends directly on the channel quality of the coexisting eMBB and URLLC users. When the channel conditions are favorable, less resources are needed for serving the eMBB and URLLC users simultaneously. Precisely, when the channel gains of the URLLC users are high, the number of frequency resources that are required to achieve the target reliability and latency and will be punctured from the ones of the eMBB users will reduce. Therefore, the data rate losses of the eMBB users will reduce as well. Alternatively, when the channel gains of the eMBB users are high, the number of frequency resources required to achieve their target data rates will reduce, and hence more resources can be punctured in order to accommodate the URLLC traffic. Therefore, the number of URLLC users that can be served jointly with the eMBB users will increase. Subsequently, the following question arises: \textit{how can one increase the channel gains of the coexisting eMBB and URLLC users? In other words, how can one enhance the propagation environment and the channel conditions of the coexisting eMBB and URLLC users?} Reconfigurable intelligent surfaces (RIS) technology has emerged as a key solution that provides answers to the above questions.\\
\indent RIS is a promising technology for next generation wireless networks that has been lately receiving a significant interest from both academia and industry due to its capability in controlling and configuring the wireless propagation environment \cite{Marco}. RIS is a planar array that is composed of a large number of passive and low-cost reflecting elements, where each can be tuned independently to a certain phase-shift \cite{9122596,8796365}. By appropriately tuning the phase-shift of each passive element, the reflected signals by the RIS can be constructively added at the points of interests \cite{9371415,8796365}. Therefore, the channel gains and the received signal strengths at the end users, which are the eMBB and the URLLC users in the context of this paper, are enhanced \cite{9122596,8796365}. As opposed to traditional relaying techniques (e.g., amplify-and-forward, and decode-and-forward), RIS exhibits a multitude of advantages. In fact, RIS offers a low cost solution that is both energy and spectral efficient \cite{9122596,8796365}. In addition, RIS is capable of passively reflecting the incident signals without additional radio frequency chains, which results in a lower power consumption. Moreover, the radio signals reflected by the RIS are free from noise corruption \cite{8796365}. Consequently, and motivated by the aforementioned benefits of RIS, this paper considers an RIS-assisted wireless network and study the problem of coexistence of services with heterogeneous requirements, namely URLLC and eMBB. The paper explores the added value of this new degree of freedom offered by the RIS technology on the performance of superimposing the URLLC traffic on a network designed to serve the eMBB users, which to the best of our knowledge, has not been explored in earlier work.}
\subsection{Related Works}
Supporting URLLC services has received considerable research interest in the last few years \cite{joint,8403963,8612914}. Particularly, URLLC applications, such as autonomous driving, factory automation and virtual reality, all need to be scheduled immediately upon arrival, and the availability of spectral resources to satisfy the strict latency and reliability requirements. On-air resource allocation for URLLC data has been shown to be more spectrally effective than other emerging techniques, such as the network slicing \cite{joint}. To enable on-air allocation, superposition and puncturing schemes have been proposed by the Third Generation Partnership Project (3GPP) standard \cite{joint}. These schemes aim to satisfy the required URLLC latency by eliminating the queuing delay. Specifically, the time domain is divided into slots where different eMBB loads can share the available frequency resources. Then, each time slot is further divided into mini-time slots for the sake of immediately serving the URLLC traffic. In fact, the arriving URLLC packets are transmitted in the following mini-slot by either puncturing or superposing onto frequency resources that are already allocated to the eMBB traffic. \\
\indent Using superposition and puncturing schemes, several works already considered the allocation of the URLLC traffic over the eMBB traffic \cite{joint,8746407,8643428,alsenwi2020intelligent,9187217,al2020downlink}. Joint scheduling of eMBB and URLLC traffic was studied in \cite{joint}, where the authors studied linear, convex and threshold models for the eMBB rate loss resulting from the superposition/puncturing scheme. The overhead associated with the URLLC load segmentation was considered \cite{8746407}, where a puncturing-based resource allocation policy was formulated with the objective of maximizing the rate utility. In \cite{8643428}, a risk-sensitive approach was introduced to alleviate the puncturing effects on eMBB users having low data rates. In \cite{alsenwi2020intelligent}, a deep reinforcement learning approach was proposed to allocate the URLLC traffic. A null-space-based spatial puncturing scheduler for joint URLLC and eMBB traffic was proposed in \cite{9187217}. A downlink puncturing scheme based on finding the similarities between the eMBB and URLLC symbols was proposed in \cite{al2020downlink}. The works of \cite{8643428,alsenwi2020intelligent,8746407,9187217,al2020downlink} consider only the puncturing scheme for joint scheduling of eMBB and URLLC loads. Authors in \cite{manzoor2020contract} have formulated a URLLC traffic allocation problem by adopting a superposition or puncturing scheme. \\
\indent Motivated by the great benefits offered by RIS in controlling and configuring the wireless propagation environments, several works studied the configuration of the phase-shifts of the RIS elements, also known as passive beamforming, to enhance the performance of wireless cellular systems \cite{8811733,8855810,9120476,cao2021ai}. The authors in \cite{8811733} proposed an alternating algorithm for passive and active beamforming design to minimize the total transmit power, where a semi-definite relaxation (SDR) approach was adopted to configure the passive beamforming. Yu et al. exploited in \cite{8855810} the fixed point iteration and manifold optimization methods to maximize the spectral efficiency in RIS-aided cellular networks. In \cite{9120476}, the RIS-aided non-orthogonal-multiple-access (NOMA) systems was studied with the objective max-min rate problem by jointly optimizing the power allocation and the RIS phase-shift matrix.
\textcolor{black}{In \cite{cao2021ai}, multi-user communications aided by single or multiple RISs were studied for multiple-input single-output (MISO) and multiple-input multiple-output (MIMO) systems. From a medium access control (MAC) perspective, the authors also came up with different possible solutions for the considered RIS-aided cellular systems}. However, a few works investigated RIS aided URLLC traffic \cite{ghanem2020joint,9206550,cao2021reconfigurable,9083788}. Authors in \cite{ghanem2020joint} proposed joint active beamforming and phase-shift optimization to allocate URLLC traffic with the objective of maximizing the URLLC sum rate in MISO system aided by RIS, in which a set of BSs cooperate to serve the URLLC traffic. The integration between the unmanned aerial vehicle (UAVs) and the RIS was studied in \cite{9206550} to support the URLLC traffic. \textcolor{black}{In \cite{cao2021reconfigurable}, the coverage and link performance of the RIS-assisted UAV systems was studied by proposing an adaptive RIS-assisted transmission protocol to control the RIS association and the RIS phase-shifts configuration. The authors also proposed a multi-task learning to reduce the time complexity of the proposed transmission protocol.} In \cite{9083788}, a grant-free access scheme aided by an RIS was proposed to enhance the URLLC reliability. \textcolor{black}{Although, the works in \cite{ghanem2020joint,9206550,cao2021reconfigurable,9083788} studied the performance of RIS-aided URLLC traffic, they considered a simple scenario where the BS serves only URLLC users. In other words, the coexistence of URLLC and eMBB traffic aided by RIS was not addressed in \cite{ghanem2020joint,9206550,cao2021reconfigurable,9083788}. Similarly, the works in \cite{8811733,8855810,9120476,cao2021ai} considered the RIS-aided wireless networks for only the eMBB service class. Hence, there is a noticeable lack in studying the coexistence problem of eMBB and URLLC traffic in RIS-aided wireless networks, which is the focus of this paper.}
\subsection{Contributions and Outcomes}
We consider an RIS-assisted cellular network that supports coexistent eMBB and URLLC services while taking into consideration the trade-off between the services regiments. In this context, we first formulate a problem for allocating the eMBB users at the beginning of each time slot. The problem aims at maximizing the eMBB sum rate with resource over-provisioning to accommodate future arrival of URLLC packets while satisfying required QoS of the different eMBB users by jointly optimizing the power allocation policy and the RIS phase-shift matrix. However, due to the coupling between the power allocation at the BS and the passive beamforming at the RIS, the formulated eMBB allocation problem is not convex, and hence, difficult to be solved. To overcome this issue, the problem is solved by applying the alternating optimization (AO) technique. Particularly, the original problem is decomposed into two sub-problems, namely, a power allocation sub-problem and a phase-shift matrix optimization sub-problem. The power allocation sub-problem is a convex problem which is solved by applying the Karush-Kuhn-Tucker (KKT) method. Meanwhile, the SDR and Gauss randomization methods are adopted to solve the phase-shift sub-problem. \\
\indent At each mini-time slot, the incoming URLLC packets should be served and transmitted simultaneously over the ongoing eMBB data using the puncturing scheme in order to satisfy the URLLC latency requirement. Hence, with the goal of maximizing the number of served URLLC packets at each mini-time slot while satisfying the eMBB and URLLC rate requirements, the frequency and power allocation should be jointly optimized along with the RIS phase-shift matrix. However, the task of optimizing the RIS configuration has a high computational time compared to the mini-time slot duration, which may violate the URLLC latency requirement. To overcome this issue, multiple RIS configurations are proactively designed and communicated to the RIS controller at the beginning of each time slot prior to the arrival of the URLLC load. Then, a control signal, if needed, is sent to the RIS controller to switch between these configurations at each mini-time slot. Once the RIS configuration is fixed, the URLLC allocation problem is formulated as a mixed integer non linear program which is difficult to be solved in a polynomial time within each mini-time slot. Hence, we convexify the problem by applying a variable change approach. Due to the nature of the URLLC service which requires high reliability, the problem is decomposed into two sub-problems 1) the URLLC admission problem and 2) the URLLC allocation problem. The first problem aims at maximizing the admitted URLLC packets while the latter aims at minimizing the eMBB loss according to the URLLC allocation strategies.\\
\indent The performance of the proposed scheme is illustrated through extensive simulations. The obtained results show that the proposed scheme achieves considerable gain on the URLLC packet admission rate and the eMBB sum rate compared to when no RIS is deployed. Moreover, the proposed switching scheme between the proactively configured RIS phase-shift matrices enhances the performance of both URLLC and eMBB services compared to the case when a fixed phase-shift matrix is used. Finally, the best allocation for the RIS is demonstrated to be close to the BS when the eMBB and URLLC users are distributed randomly on the coverage area of the BS.
\subsection{Paper Outline}
The rest of the paper is organized as follows. Section \ref{sec:system_model} presents the system model. Section \ref{problem formulation} presents the problem formulations for the eMBB and URLLC allocation. Section \ref{Roadmap} presents the roadmap for the solution approach. Sections \ref{urllc_sec} presents the proposed solution approach for the URLLC allocation problem. Sections \ref{Simulation results} and \ref{Conclusion} present the simulation results and the conclusion, respectively.
\section{System Model and Rate Analysis}
\label{System Model}
\subsection{System Model}
\label{sec:system_model}
\begin{table}
\begin{minipage}{0.5\linewidth}
\centering
\includegraphics[width=1\columnwidth,draft=false]{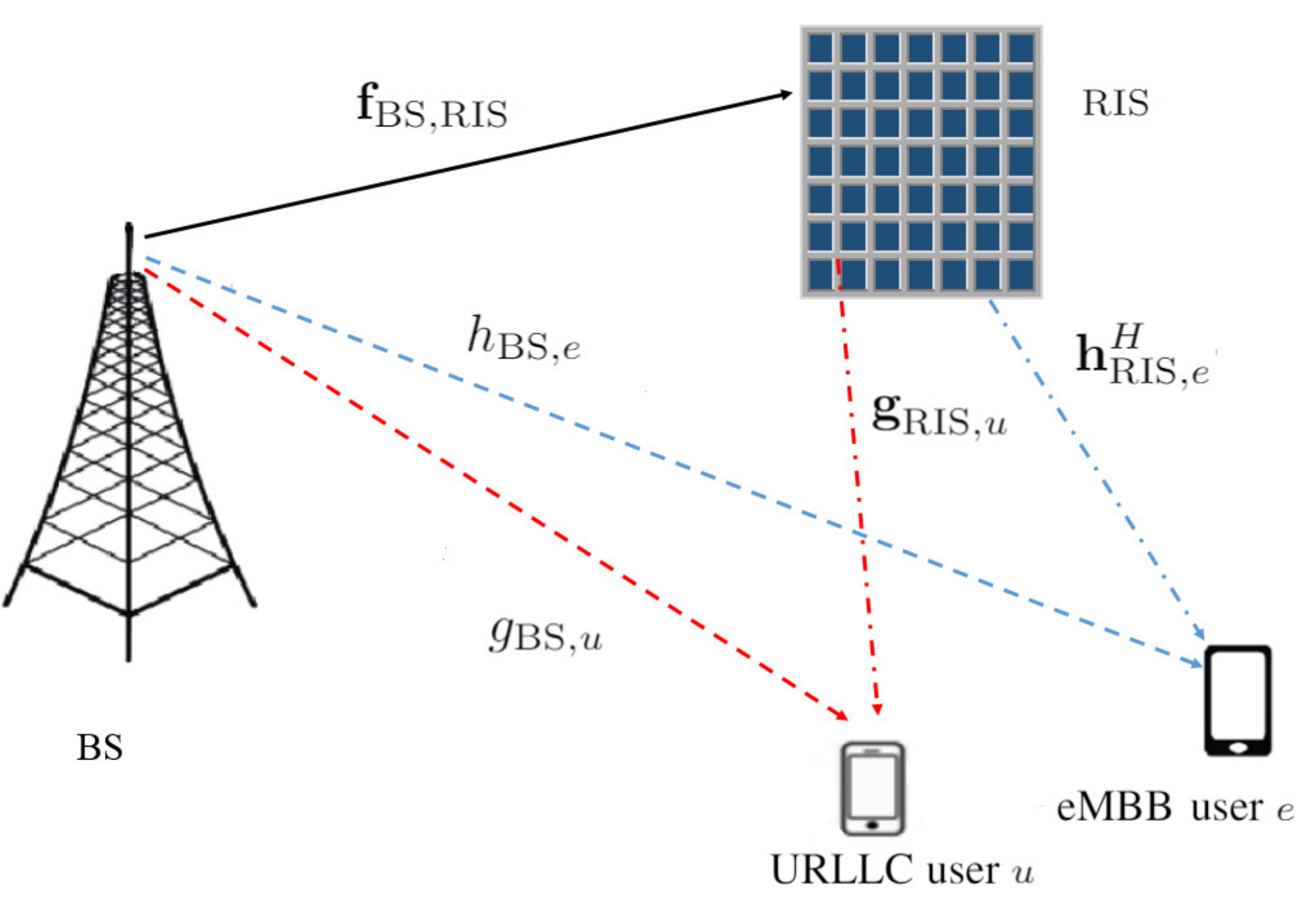}
\captionof{figure}{System Model.}
\label{Fig:System_Model}
\end{minipage}
\hspace{0.5cm}
\begin{minipage}{0.5\linewidth}
\caption{Table of symbols used in the paper.}
\renewcommand{\arraystretch}{.3} 
\setlength{\tabcolsep}{0.05cm} 
\centering 
\begin{tabular}{|c|c|}
    \hline
    Symbol& Description   \\
    \hline
    $\mathcal{E}$& Set of eMBB users\\
    \hline
    $\mathcal{U}$& Set of URLLC users\\    
    \hline
    $\mathcal{N}$& Set of RIS elements \\    
    \hline
    $\mathcal{M}$& Set of mini-time slots  \\    
    \hline
    \multirow{2}{*}{$\mathcal{L}^m$}& Set of URLLC packets at mini-time \\ 
    &  slot $m$\\
    \hline
    $E$& Number of eMBB users\\
    \hline
    $U$& Number of URLLC users\\
    \hline
    $N$& Number of RIS elements \\
    \hline
    \multirow{2}{*}{$M$}& Number of mini-time slots per time \\ 
    &  slot  \\   
    \hline
    \multirow{2}{*}{$L^m$}& Number of URLLC packets  \\   
    & at mini-time slot $m$ \\
    \hline
    $P_{\rm BS}$& Power budget at the BS\\
    \hline
    $\boldsymbol{\Phi}$& RIS phase-shift matrix\\
    \hline
    $B$& Number of resource blocks  \\
    \hline
    $\tau$ & Mini-time slot duration  \\
    \hline
    $W$& Resource block bandwidth  \\
    \hline
    $r_{\rm th}$& eMBB rate threshold \\
    \hline
    $c_{\rm th}$& URLLC rate threshold \\
    \hline
\end{tabular}
\label{Table:symbols} 
\end{minipage}
\vspace{-0.3in}
\end{table}
\indent We consider a downlink radio access network consisting of a single base station (BS), equipped with one single antenna, serving several spatially dispersed eMBB and URLLC users, each equipped with one single antenna.\footnote{\textcolor{black}{In this work, we consider the case of single antenna BS and cellular users. The main motivation behind such a choice is that a proof of concept for multiplexing URLLC and eMBB traffic with the aid of RIS is aimed to be investigated in a simple setup, so that the fundamental insights and observations can be obtained. Nevertheless, the same analysis can be extended for the case when the BS and/or the cellular users are equipped with multiple antennas.}} The total bandwidth allocated to the BS is divided into $B$ resource blocks (RBs), each of bandwidth $W$. The service period of cellular users is divided into equally sized time slots. Each time slot is further divided into a set of $M$ equally sized mini-time slots, denoted by $\mathcal{M}\triangleq\{1,2,\dots,M \}$, where the duration of each mini-time slot is denoted by $\tau$. Let $\mathcal{E}\triangleq\{1,2,\dots,E \}$ and  $\mathcal{U}\triangleq\{1,2,\dots,U \}$ denote the sets of the eMBB and URLLC users, respectively, where $E$ and $U$ denote the total numbers of eMBB and URLLC users, respectively, that are simultaneously communicating with the BS within one time slot. The eMBB users are admitted at the beginning of each time-slot and the adopted multiple access technique is the orthogonal frequency-division-multiple-access (OFDMA). In addition, the eMBB users share equally the available frequency resources, i.e., each eMBB user has $b = \frac{B}{E}$ RBs. The URLLC load, on the other hand, can arrive within the serving time slot, i.e., during one mini-time slot within the same time slot, and it should be served immediately to satisfy its latency requirements. Hence, the URLLC packets are immediately transmitted upon arrival in the following mini-time slot by puncturing the frequency resources that are already allocated to the eMBB users at the beginning of the time slot. Moreover, for all $u \in \mathcal{U} $, the arrival process of URLLC packets per mini-time slot of the $u$th URLLC user is assumed to follow a Poisson distribution with an average arrival rate $\lambda_u$ \cite{9020161}. The symbols used throughout the paper are listed in Table \ref{Table:symbols}. \\
\indent One single RIS, equipped with $N$ reflective elements, is deployed to dynamically control the propagation environment between the BS and the different eMBB and URLLC users. Each reflective element consists of an atom that can adjust the phase of each incident wave. Let $\mathcal{N}$ denote the set of indices between $1$ and $N$, i.e., $\mathcal{N}\triangleq\{1,2,\dots,N \}$. Then, we denote by $\boldsymbol{\Phi} \triangleq {\rm diag} \left(e^{j\phi_1},e^{j\phi_2},...,e^{j\phi_N} \right) \in \mathbb{C}^{N \times N}$ the phase-shift matrix of the RIS, where for all $n \in \mathcal{N}$, $\phi_n\in[0,2\pi)$ denotes the phase-shift of the $n$th RIS reflective element. The RIS is connected to a control unit that adjusts the phase-shift matrix $\boldsymbol{\Phi}$. The channel state information (CSI) of all communicating nodes, including the eMBB and URLLC users and the RIS, are assumed to be perfectly known at the BS \cite{8811733}.\footnote{We assume that the BS knows the locations of the RIS and of all cellular users, which can be used then to perfectly estimate the different communication nodes at the beginning of each time slot.} In this context, let $\mathbf{f}_{{\rm  BS},{\rm  RIS}}\in\mathbb{C}^{N\times1}$ denote the vector that contains the channel coefficients between the BS and the RIS elements. Moreover, for all $e \in \mathcal{E}$, let $h_{{\rm  BS},e}\in\mathbb{C}$ and $\mathbf{h}_{{\rm RIS}, e}\in\mathbb{C}^{N\times1}$ denote the channel coefficients from the BS to the $e$th eMBB user and from the RIS to the $e$th eMBB user, respectively. Additionally, for all $u \in \mathcal{U}$, let $g_{{\rm BS},u}\in\mathbb{C}$ and $\mathbf{g}_{{\rm RIS}, u}\in\mathbb{C}^{N\times1}$ denote the channel coefficients from the BS to the $u$th URLLC user and from the RIS to the $u$th URLLC user, respectively. Each communication link in the network is assumed to have a quasi-static flat-fading Rayleigh channel, except the ones between the BS and the RIS elements, which are assumed to have a Rician channel model. This is basically due to the fact that a necessary condition for the deployment of RIS in any cellular system is that the BS has a direct line-of-sight with the RIS.
\vspace{-0.2in}
\subsection{Signal Model and Rate Analysis}
Considering both the direct link and the cascaded link through the RIS between the BS and each eMBB user, the received signal per RB at the $e$th eMBB user, for all $e \in \mathcal{E}$, can be expressed as  \cite{9279247}
\begin{equation}
    y_e = \left( h_{{\rm BS}, e} + \mathbf{h}_{{\rm RIS},e}^H \boldsymbol{\Phi} \,\mathbf{f}_{{\rm  BS},{\rm  RIS}} \right) \sqrt{p_e}\,x_e+z_e,
\end{equation}
where $x_e$ is the signal that contains the data of the $e$th eMBB user to be transmitted throughout the entire time-slot, $p_e$ is its associated allocated power per RB at the beginning of the time slot and $z_e$ is the additive white Gaussian noise (AWGN) experienced at the $e$th eMBB user throughout the entire time slot, which is assumed to be $\mathcal{CN}(0,\sigma^2)$ distributed. \textcolor{black}{Accordingly, by considering a target block error rate (BLER) $\epsilon_{\rm eMBB}$ for all eMBB users, the data rate in [bits/s] of the $e$th eMBB user per RB, for all $e \in \mathcal{E} $, can be expressed as
\begin{equation}
\tag{3}
  r_e \left(\boldsymbol{\Phi}, p_e \right)=W\,\log_2 \left(1+\frac{p_e|h_{{\rm BS}, e}+  \mathbf{h}_{{\rm RIS},e}^H\boldsymbol{\Phi} \,\mathbf{f}_{{\rm  BS},{\rm  RIS}}|^2}{\Gamma_{\rm eMBB}\sigma^2}\right),
\end{equation}
where $\Gamma_{\rm eMBB}=\frac{-\ln \left(5\, \epsilon_{\rm eMBB}\right)}{0.45}$ represents the SNR gap between the Shannon capacity and the achievable rate of the adopted modulation scheme, when the target BLER is $\epsilon_{\rm eMBB}$ \cite{9020161}.} \\ 
\indent For $m \in \mathcal{M}$, let $\mathcal{L}^m\triangleq\{1,2,\dots, L^m\} $ denote the set of the arrived URLLC packets that need to be transmitted at mini-time slot $m$, where $L^m$ is the number of URLLC packets at mini-time slot $m$. Then, for all $m \in \mathcal{M}$ and $l \in \mathcal{L}^m$, the received signal at the $l$th URLLC packet during the $m$th mini-time slot can be expressed as
\begin{equation}
    y_l^{m} = \left(g_{{\rm BS}, l}+  \mathbf{g}_{{\rm RIS},l}^H \boldsymbol{\Phi} \,\mathbf{f}_{{\rm  BS},{\rm  RIS}} \right) \sqrt{p_l^m}\,x_l^m+z_{l}^m,  
\end{equation}
where $x_l^m$ is the signal that contains the data of the $l$th URLLC packet to be transmitted within the $m$th mini-time slot, $p_l^m$ is its associated allocated power and $z_{l}^m$ is the AWGN experienced at the URLLC user associated to the $l$th URLLC packet within the $m$th mini-time slot, which is assumed to be $\mathcal{CN}(0,\sigma^2)$ distributed. Moreover, $g_{{\rm BS}, l}\in \{g_{{\rm BS}, 1}, g_{{\rm BS}, 2}, \dots, g_{{\rm BS}, U}\} $ and $\mathbf{g}_{{\rm RIS},l}\in \{\mathbf{g}_{{\rm RIS},1}, \mathbf{g}_{{\rm RIS},2}, \dots, \mathbf{g}_{{\rm RIS},U}\} $ are the channel coefficients from the BS and from the RIS to the URLLC user associated to the $l$th URLLC packet, respectively. Consequently, by considering a target BLER $\epsilon_{\rm URLLC}$ for all URLLC users, the achievable rate per RB of the $l$th URLLC packet, for all $l \in \mathcal{L}^m$, within the $m$th mini-time slot, can be expressed as
\begin{equation}
  c_{l} \left(\boldsymbol{\Phi}, p_l^m\right)=  W\,\log_2\left(1+\frac{p_l^m|g_{{\rm BS}, l}+  \mathbf{g}_{{ \rm RIS},l}^H \boldsymbol{\Phi} \,\mathbf{f}_{{\rm  BS},{\rm  RIS}}|^2}{\Gamma_{\rm URLLC}\sigma^2}\right), 
\end{equation}
where $\Gamma_{\rm URLLC}=\frac{-\ln \left(5\, \epsilon_{\rm URLLC}\right)}{1.25}$ represents the SNR gap between the Shannon capacity and the achievable rate of the adopted modulation scheme, when the target BLER is $\epsilon_{\rm URLLC}$ \cite{9020161}.
\section{Problem formulation and Methodology}
\label{problem formulation}
\subsection{eMBB Allocation}
We aim to design an effective puncturing scheme to multiplex the URLLC traffic over the existing eMBB traffic in an RIS-aided cellular network. The objective is to enhance the system performance in terms of the eMBB throughput and to guarantee simultaneously the different requirements of both eMBB and URLLC services. At the beginning of each time slot, the BS allocates its budget of power and designs the RIS phase-shift matrix in order to serve the eMBB users.\footnote{\textcolor{black}{The eMBB power allocation and RIS phase-shift matrix optimization can be done for one or more eMBB time-slots, depending on the variations of the CSI of the eMBB and URLLC users \cite{zhang2020millimeter}.}} With the objective of maximizing the eMBB sum rate subject to a QoS constraint for each eMBB user and a budget power constraint at the BS, the joint power allocation and RIS phase-shift matrix design can be given by the following optimization problem.
\allowdisplaybreaks
\begingroup
\begin{subequations}
\begin{align}
  \mathcal{P}_1: \quad &\max_{\boldsymbol{\Phi},\mathbf{p}_{\rm e}}\,\, \sum_{e=1}^E r_e\left(\boldsymbol{\Phi},{p_e}\right)\label{P1_obj}\\
   &\text{s.t.} \quad  \left(1-\delta\right)b\,r_e(\boldsymbol{\Phi},{p_e})\ge {r_{\rm th}},\quad \forall\,\, e \in \mathcal{E},\label{P1_C1}\\
   &\qquad \sum_{e=1}^E b\,p_e\le P_{\rm BS},\label{P1_C2}\\
   &\qquad 0\le\phi_n< 2\pi, \quad \forall\,\, n \in \mathcal{N},\label{P1_C3}
\end{align}
\end{subequations}
\endgroup
where $\mathbf{p}_{\rm e} = \left[p_1,p_2,...,p_E \right]^T$, $P_{\rm BS}$ is the total power of the BS and $\delta \in [0,1]$ is a predefined rate margin factor. Constraint \eqref{P1_C1} can be rewritten, for all $e \in \mathcal{E}$, as 
\begin{equation}
    r_e(\boldsymbol{\Phi},{p_e}) \ge \frac{r_{\rm th}}{b} + \delta r_e(\boldsymbol{\Phi},{p_e}), 
\end{equation}
where $r_e(\boldsymbol{\Phi},{p_e})$ is the achievable rate of the $e$th eMBB user, which should be guaranteed even when some of its allocated RBs are punctured within the time-slot, $r_{\rm th}$ is the minimum required data rate per eMBB user, and $\delta r_e(\boldsymbol{\Phi},{p_e})$ depicts a rate surplus as a result of over provisioning of resources for the URLLC load at the beginning of the time slot. This over-provisioned resources will be of utility for URLLC users to use at the time of arrival of URLLC packets. In other words, some RBs of the $e$th eMBB user will be punctured within the time slot as long as the resulting rate-loss does not exceed $\delta r_e(\boldsymbol{\Phi},{p_e})$. Additionally, constraint \eqref{P1_C2} guarantees that the total transmit power by the BS does not exceed its power budget. Problem $\mathcal{P}_1$ will be solved at the beginning of the time slot within which the BS will transmit the data of the $E$ eMBB users. Afterwards, the BS will keep employing the obtained optimal power allocation scheme $\mathbf{p}_{\rm e}^*$ and the optimal RIS phase-shift matrix $\boldsymbol{\Phi}_{\rm e}^*$ over the following time slots as long as the number of eMBB users $E$ and the CSI of all eMBB users do not change.\\ 
\indent We note that, owing to constraint \eqref{P1_C1}, problem $\mathcal{P}_1$ may not be always feasible. Hence, if the problem is not feasible for the eMBB users, we solve an eMBB admission problem to guarantee the feasibility conditions of problem $\mathcal{P}_1$. The admission problem aims to select a set of eMBB users $\mathcal{E}_f\triangleq\{1,2\dots,E_f\}$, where $E_f\le E$, based on their contribution on the eMBB sum rate such that $\mathcal{P}_1$ is feasible. To do this, we first solve problem $\mathcal{P}_1$. If the problem is not feasible, we resolve the same problem while setting $r_{\rm th}=0$. Then, we remove the eMBB user with the lowest contribution on the eMBB sum rate. These steps are repeated until a set of eMBB users $\mathcal{E}_f  \subseteq  \mathcal{E}$ can be admitted.
\subsection{URLLC Allocation}
\indent Within one time-slot, the URLLC packets are allocated within any mini-time slot upon arrival in order to satisfy their latency requirement. Such allocation is performed by puncturing the frequency RBs of the eMBB users and distributing them over the different URLLC packets to be transmitted. Within this scheme, the reliability of the URLLC load needs also to be satisfied. In fact, let us assume that, for all $m \in \mathcal{M}$, the incoming URLLC packets have the same size, which is denoted by $\zeta$ [bits]. In this case, for all $m \in \mathcal{M}$ and $l \in \mathcal{L}^m$, the $l$th URLLC packet arriving at the $m$th mini-time slot must be entirely and successfully transmitted to the relative URLLC users. Such QoS constraint can be expressed, for all $m \in \mathcal{M}$, as
\begin{equation}
\label{eq:urllc-QoS-1}
I_l^m c_l(p_l^m,\boldsymbol{{\Phi}}^m) \ge c_{\rm th},
\end{equation}
where $I_l^m=\sum_{e=1}^E I_{e,l}^m$, in which for all $e \in \mathcal{E}_f$, $I_{e,l}^m \in \{0,1,\dots,b\}$ is the number of RBs punctured for the $e$th eMBB user and allocated to the $l$th URLLC packet, $\boldsymbol{\Phi}^m \triangleq {\rm diag} \left(e^{j\phi_1^m},e^{j\phi_2^m},...,e^{j\phi_N^m} \right) \in \mathbb{C}^{N \times N}$ is the RIS phase-shift matrix at mini-time slot $m$ and $c_{\rm th} = \frac{\zeta}{\tau}$. Furthermore, puncturing the RBs of any eMBB user may violate its QoS requirement. In fact, according to the linear rate loss model \cite{joint}, the instantaneous achievable rate of the e$th$ eMBB, for all $e \in \mathcal{E}_f$, at mini-time slot $m$, for all $m \in \mathcal{M}$, is given by
\begin{equation}\label{mini_slot_rate}
  R_e\left(\boldsymbol{\Phi}^m,p_e,I_e^m\right)= \left(1\,-\frac{I_e^m}{b} \right)\,r_e\left(\boldsymbol{\Phi}^m,p_e\right),
\end{equation}
where $I_{e}^m=\sum_{l=1}^{L^m} I_{e,l}^m$ is the total number of punctured RBs from the e$th$ eMBB user at mini-time slot $m$, which must verify $I_e^m \in \{0,1,\dots,b\}$. \textcolor{black}{Afterwards, at each mini-time slot, we should ensure that puncturing the eMBB resources while admitting the URLLC packets does not impact adversely the eMBB QoS, which is represented by the minimum data rate $r_{\rm th}$ for the entire $M$ mini-time slots. To guarantee this, we assume that the URLLC scheduler at the BS is causal so it only knows the current and the past states of the URLLC load. As a result, at each mini-time slot $m$, the BS assumes that no URLLC packets will arrive at the subsequent mini-time slots $\{ m+1,\dots, M\}$, i.e. $\sum_{i=m+1}^ML^i=0$. Then, for all $m \in \mathcal{M}$ and $e \in \mathcal{E}_f$, the QoS constraint of the $e$th eMBB user at mini-time slot $m$ that ensures the eMBB QoS rate constraint for the entire time slot can be expressed as  
\begin{equation}\label{eq;embb_mini}
\frac{\sum_{i=1}^m R_e\left(\boldsymbol{\Phi}^i,{p_e}^i,I_{e}^i\right)+ \left(M-m\right) r_e(\boldsymbol{\Phi},p_e)}{M}\ge \frac{r_{\rm th}}{b\,},
\end{equation}
which can be simplified as 
\begin{equation}\label{eq:rate_cst_mini_slot}
 R_e\left(\boldsymbol{\Phi}^m,{p_e}^m,I_{e}^m\right)\ge r_{e,\rm th}^{'},
\end{equation}
where 
\begin{equation}
    r_{e,\rm th}^{'} =\frac{M\,r_{\rm th}}{b}-\left(\sum_{i=1}^{m-1}R_e\left(\boldsymbol{\Phi}^i,{p_e}^i,I_{e}^i\right)+\left(M-m\right) r_e(\boldsymbol{\Phi},p_e)\right).
\end{equation}
The inequality in \eqref{eq:rate_cst_mini_slot} guarantees that the eMBB loss at each mini-time slot, resulting from puncturing the eMBB RBs and adjusting the RIS phase-shift matrix, does not impact the eMBB rate requirements $r_{th}$ for the entire $M$ mini-time slots.} \\ 
\indent \textcolor{black}{The URLLC allocation problem aims at maximizing the admitted URLLC packets while minimizing the eMBB rate loss at each mini-time slot while guaranteeing the target QoS of the eMBB users and the rate requirements of the URLLC packets. For all $m \in \mathcal{M}$, the number of the admitted URLLC packets at the $m$th mini-time slot (which we aim to maximize) can be expressed as
\begin{equation}
    f_1(\mathbf{k}^m)=\sum_{l=1}^{L^m} k_l^m,
\end{equation}
where $\mathbf{k}^m = \left[k_1^m,k_2^m,...,k_{L^m}^m \right]^T$ is a $L^m \times 1$ binary vector that represents the admission of the URLLC packets, i.e., for all $l \in \{1,2,\dots,L^m\}$, if $k_l^m=1$, then the $l$th URLLC packet is admitted, and $k_l^m=0$, then it is not admitted. Based on this, the admission rate of the URLLC load, denoted by $\eta$, is defined as the total number of URLLC packets that are successfully served at each time slot divided by the total number of arrived URLLC packets at the same time slot, i.e., $\eta = \frac{\sum_m^M\hat{L}^m}{\sum_m^M{L}^m}$, where, for all $m \in \mathcal{M}$, $\hat{L}^m$ is the number of served URLLC packets at mini-time slot $m$.\\
\indent On the other hand, for all $m \in \mathcal{M}$, the overall eMBB traffic rate loss (which we aim to minimize) can be expressed as
\begin{align}
    \label{eq:objective_rate_loss}
   f_2(\mathbf{I}^m)= \sum_{e=1}^{E_f} I_e^m\beta_e^{\pi,m},
\end{align}
where for all $e \in \mathcal{E}$, $\beta_e^{\pi,m} \in [0,1]$ is the weight of allocating the URLLC load on the $e$th eMBB user at mini-time slot $m$, which depends on the URLLC allocation strategy, denoted by $\pi$. The objective function $f_2$ is a weighted sum of the number of frequency resources to be punctured $\left(I_e^m\right)_{1\leq e\leq E_f}$, and hence, it is a convex function. Therefore, the optimal minimization strategy of the objective function $f_2$ is the one that punctures higher number of frequency resources from the eMBB users with lower weights, i.e., for all $(i,j) \in \mathcal{E}_f$, if $\beta_{i}^{\pi,m} \leq \beta_{j}^{\pi,m}$, then $I_{i}^m \geq I_{j}^m$. At this stage, one might think on how to efficiently design weight of allocating the URLLC load at each mini-time slot $m$, for all $m \in \mathcal{M}$. In practice, several URLLC allocation strategies, such as random allocation, minimum eMBB rate loss and proportional fairness can be adopted to distribute/control the eMBB loss \cite{joint}. For all $m \in \mathcal{M}$, let $\boldsymbol{\beta}^{m,\pi} = \left[\beta_1^{m,\pi},\beta_2^{m,\pi},\dots,\beta_{E_f}^{m,\pi} \right]^T$ denote the $E_f\times 1$ vector of puncturing weights of the allocation strategy $\pi$, which can be illustrated, when the URLLC allocation strategy $\pi$ is the minimum eMBB rate loss and the proportional fairness URLLC, as follows:
\begin{enumerate}
\item \textit{Minimum eMBB Rate Loss (MeRL):} This strategy aims at minimizing the eMBB rate loss. Based on the observation above, the BS allocates lower weight to the eMBB users that are susceptible to have low rate losses, i.e., the users with low achievable data rates compared to the eMBB users with high achievable data rates. Hence, this strategy aims to minimize the eMBB rate loss by puncturing the eMBB users with low achievable rates. Thus, for all $e \in \mathcal{E}_f$, $\beta_e^{m, \pi}$ is expressed as
\begin{equation}
      \beta_e^{m,\pi}=\frac{\hat{R}_e^m}{\sum_{e=1}^E \hat{R}_e^m},
    \end{equation}
    where $\hat{R}_e^m$ is the maximum allowed rate loss for the $e$th eMBB user at mini-time slot $m$, and it is expressed as
   \begin{align} 
   \hat{R}_e^m &= r_e \left(\boldsymbol{\Phi}^m, p_e \right)-r_{e,\rm th}^{'} \nonumber\\
      &={\sum_{i=1}^{m-1}  R_e\left(\boldsymbol{\Phi}^i,p_e,I_e^i\right)+(M-m+1)\, r_e \left(\boldsymbol{\Phi}, p_e \right)}-\frac{M\times r_{th}}{b}.
    \end{align}
\item \textit{Proportional Fairness (PF):} This strategy aims at ensuring certain fairness between the eMBB users. Based on the observation above, the BS allocates lower weight to the eMBB users with high data rates compared to the ones with lower data rates. This scheme encourages the puncturing of eMBB users with higher rates. Consequently, for all $e \in \mathcal{E}_f$, the weight $\beta_e^{m,\pi}$ is expressed as
    \begin{equation}
       \beta_e^{m,\pi}=1-\frac{\hat{R}_e^m}{\sum_{e=1}^E \hat{R}_e^m}.
    \end{equation}
\end{enumerate}}
\indent \textcolor{black}{Based on the above discussion, and for all $m \in \mathcal{M}$, the URLLC allocation problem is formulated at the $m$th mini-time slot as
\allowdisplaybreaks
\begingroup
\begin{subequations}
\begin{align}
  \mathcal{P}_2^m:  \quad &\max_{\boldsymbol{\Phi}^m,\mathbf{p}_{\rm L}^m,\mathbf{k}^m , \mathbf{I}^m} [f_1(\mathbf{k}^m),\,-f_2(\mathbf{I}^m)] \label{P2_obj} \\
    &\text{s.t.}  \quad I_{l}^m\,c_l(p_l^m,\boldsymbol{\Phi}^m) \ge k_l^m c_{th}, \qquad\qquad \qquad \forall l \in \mathcal{L}^m, \label{P2_c1} \\
    &\qquad R_e\left(\boldsymbol{\Phi}^m,{p_e},I_{e}^m\right)\ge r_{e,\rm th}^{'}, \qquad \qquad \qquad \,\, \forall e\in \mathcal{E}_f, \label{P2_c2} \\
    &\qquad \sum_{e=1}^{E_f} (b- I_{e}^m) p_e+\sum_{l=1}^{L^m} p_l^m I_{l}^m\le P_{\rm BS}, \label{P2_c3} \\
    &\qquad  k_{l}^m \in\{0,1\}, \qquad \qquad \qquad \qquad \qquad \,\,\,\,\, \forall l\in \mathcal{L}^m,\label{P2_c4}\\
    &\qquad\frac{I_l^m}{B}\le k_{l}^m\le I_l^m \qquad \qquad \qquad \qquad \quad\,\,\,\, \forall l\in \mathcal{L}^m,\label{P2_c5}\\
    &\qquad  I_{e, l}^m \in \{0,1,\dots,b\},  \qquad \qquad\qquad \quad \quad \, \forall e\in \mathcal{E}_f, \forall l\in \mathcal{L}^m, \label{P2_c6} \\
    &\qquad 0\le\phi_n< 2\pi,\qquad \qquad \qquad \qquad \qquad \,\,\, \forall n\in \mathcal{N},\label{P2_c7} 
\end{align}
\end{subequations}
\endgroup
where $\mathbf{I}^m = \left(I_{e,l} \right)_{\underset{1 \leq l \leq L^m}{1\leq e\leq E_f}}$ and $\mathbf{p}_{\rm L}^m = \left[p_1^m,p_2^m,...,p_{L^m}^m \right]^T$. Based on this formulation, and for all $m \in \mathcal{M}$, the URLLC allocation problem at mini-time slot $m$ consists of obtaining the optimal power and frequency resource allocation for the URLLC packets, which are represented by $\mathbf{p}_{\rm L}^m$ and $\mathbf{I}^m$, respectively, along with the optimal RIS phase-shift matrix $\boldsymbol{\Phi}^m$ at each mini-time slot $m$}. Constraint \eqref{P2_c1} represents the QoS requirement of the URLLC packets. The eMBB QoS constraints are guaranteed in \eqref{P2_c2}. Constraint \eqref{P2_c3} indicates that the allocated eMBB and URLLC power should not exceed the total BS power. For all $l \in \mathcal{L}^m$, constraint \eqref{P2_c4} indicates that the admission variable $k_l$ is binary, whereas constraint \eqref{P2_c5} guarantees that, if the $l$th URLLC packet is allocated, i.e., $k_l^m=1$, then its allocated resources must be greater than zero, i.e., ${I}^m_l>0$. Constraint\eqref{P2_c6} indicates that, for all $e \in \mathcal{E}_f$, the punctured resources of the $e$th eMBB user should not exceed its total frequency resources $b$. 
\section{The Solution Roadmap}\label{Roadmap}
\textcolor{black}{
\indent As illustrated in the previous section, the eMBB allocation problem $\mathcal{P}_1$ aims at maximizing the eMBB sum rate over an entire time slot, whereas for all $m \in \mathcal{M}$ the URLLC allocation problem $\mathcal{P}_2^m$ aims at jointly maximizing the number of admitted  URLLC packets and minimizing the eMBB rate loss at each mini-time slot $m$, while maintaining the QoS of the URLLC and eMBB services. For all $m \in \mathcal{M}$, the URLLC allocation problem $\mathcal{P}_2^m$ is a mini-time slot basis problem which should be solved whenever the URLLC load exists. Although optimizing the RIS phase-shift matrix at each mini-time slot is going to provide optimal solution for both eMBB and URLLC traffic, this approach may impact the URLLC latency requirements. Specifically, for all $m \in \mathcal{M}$, problem $\mathcal{P}_2^m$ is a very complex problem, and it should be solved within less than one millisecond. Now, although sub-optimal solutions for problem $\mathcal{P}_2^m$ may be attained using iterative methods for all $m \in \mathcal{M}$, the iterative methods usually need high computational time which could exceed the URLLC latency constraint. }
\subsection{Methodology}
\label{Methodology}
\begin{figure}[t]
	\centering
	\center{\includegraphics[width=.8\columnwidth,draft=false]{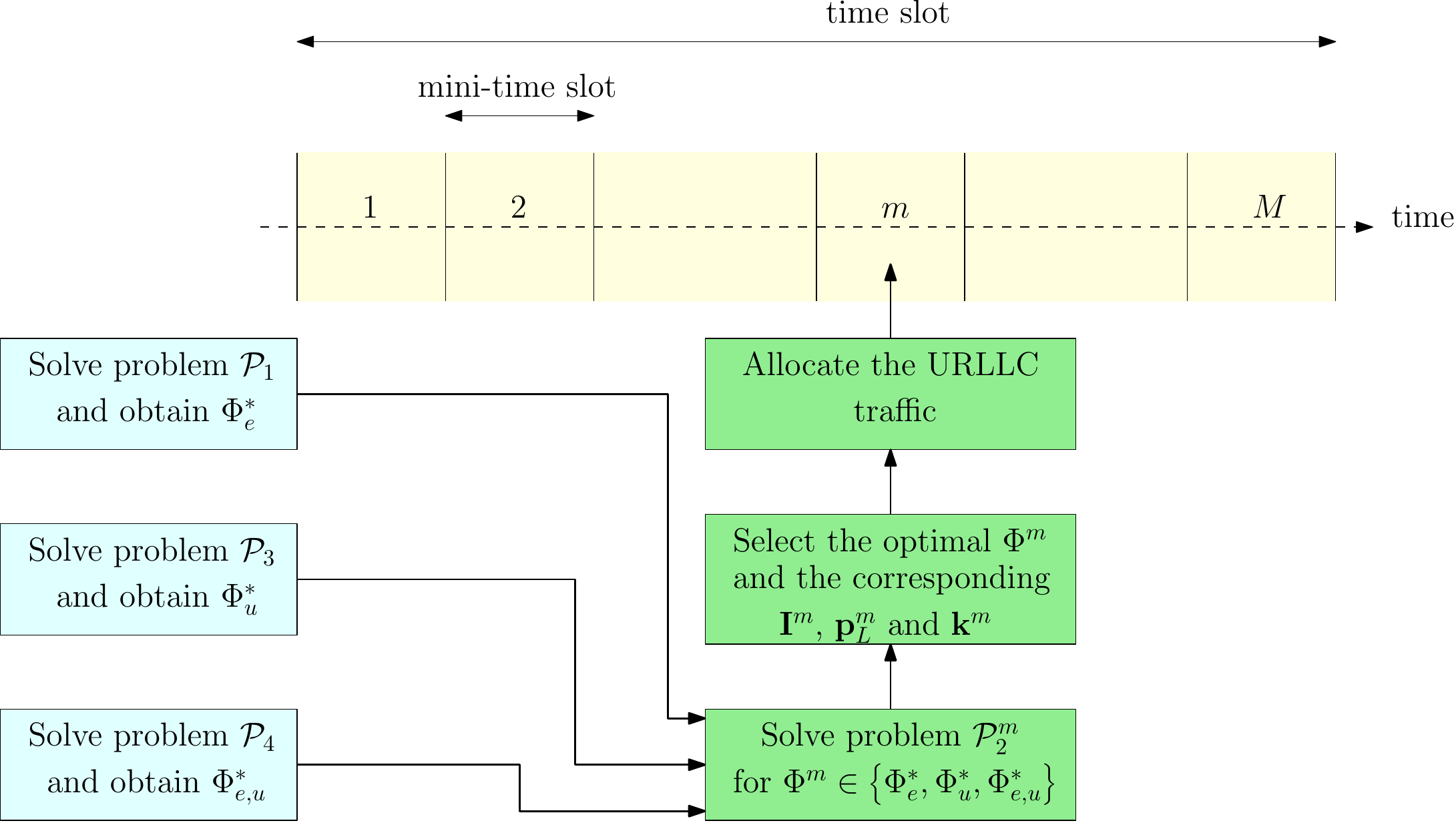}}
	\caption{Proposed Methodology }
	\label{Fig:Proposed_Methodology}
	\vspace{-0.3in}
\end{figure}
\textcolor{black}{In this section, we propose an efficient approach to alleviate the high computational complexity of optimizing the RIS phase-shift matrix per mini-time slot. The main idea is to move the RIS phase-shift matrix optimization, that will be used to jointly serve the eMBB and URLLC services at each mini-time slot, at the beginning of the associated time-slot. In other words, as shown in Fig. \ref{Fig:Proposed_Methodology}, the RIS phase-shift matrix will be proactively designed at the beginning of the time-slot in a way that can possibly satisfy the requirements of both the existing eMBB traffic and the upcoming URLLC traffic at each mini-time slot. Hence, the pre-configured RIS phase-shift matrix can be employed directly at each mini-time slot whenever a URLLC traffic is present. Accordingly, the BS just sends control signals to the RIS to switch between the pre-computed phase-shift matrices. Now, the question that arises here is the following. \textit{"Without prior knowledge of the upcoming URLLC traffic, based on which criteria the RIS phase-shift matrix will be optimized?"} To answer this question, we propose in the following three different approaches that can be used.} 
\subsection{eMBB RIS phase-shift matrix $\boldsymbol{\Phi}_{\rm e}^*$} 
This approach consists of using the optimal phase-shift matrix $\boldsymbol{\Phi}_{\rm e}^*$ that is obtained from solving the eMBB allocation problem $\mathcal{P}_1$ at each mini-time slot, even when the URLLC traffic does exist. However, it is not straightforward to solve problem $\mathcal{P}_1$ directly due to the non-convexity of its objective and and constraints, as well as the high coupling between the transmit power and the phase-shift matrix. With the aid of alternating optimization (AO), problem $\mathcal{P}_1$ is decomposed into two sub-problems, a power allocation sub-problem and an RIS phase-shift matrix optimization sub-problem, which are solved alternately \cite{elhattab2021reconfigurable}. These two sub-problems are detailed next.\\
\indent For a fixed RIS phase-shift matrix, problem $\mathcal{P}_1$ is reduced to a power allocation problem that can be formulated as shown in problem $\mathcal{P}_{1,1}$ in \eqref{PowerAllocation} on top of this page. Based on this formulation, problem $\mathcal{P}_{1,1}$ is a convex optimization problem that can be easily solved by applying the KKT condition. On the other hand, for a feasible power solution $\mathbf{p}_{\rm e}$, the RIS phase-shift optimization sub-problem is written as shown in problem $\mathcal{P}_{1,2}$ in \eqref{phaseshiftoptimization} on top of this page. Due to non convexity of its objective function and its constraints, problem $\mathcal{P}_{1,2}$ is a non-convex problem. In order to tackle this challenge, the SDR technique along with the Gaussian randomization are applied \cite{elhattab2021reconfigurable}. The details of the solution approach of problem  $\mathcal{P}_{1,2}$ are provided in Appendix A.
\begin{figure*}
\begin{minipage}{.5\textwidth}
\begin{subequations}
\label{PowerAllocation}
\begin{align}
  {\mathcal{P}}_{1,1}:\quad & \max_{\mathbf{p}_{\rm e}}\,\, \sum_{e=1}^{E_f} r_e\left(\boldsymbol{\Phi},{  p_{ e}}\right)\label{P1_obj_A}\\
   \text{s.t.}\quad  & \left(1-\delta\right)\,b\,r_e(\boldsymbol{\Phi},{  p_{ e}})\ge {r_{\rm th}},\quad \forall\,\, e \in \mathcal{E}_f,\label{P1_C1_A}\\
   &\qquad \sum_{e=1}^{E_f} b\,p_e\le P_{\rm BS}.\label{P1_C2_A}
\end{align}
\end{subequations}
\end{minipage}
\begin{minipage}{.5\textwidth}
\begin{subequations}
\label{phaseshiftoptimization}
\begin{align}
  {\mathcal{P}}_{1,2}: \quad &\max_{\boldsymbol{\Phi}}\,\, \sum_{e=1}^{E_f} r_e\left(\boldsymbol{\Phi},{  p_{ e}}\right)\label{PS1_obj}\\
  \text{s.t.} \quad & \left(1-\delta\right)\,b\,r_e(\boldsymbol{\Phi},{  p_{ e}})\ge {r_{\rm th}},\quad \forall\,\, e \in \mathcal{E}_f,\label{PS1_C1}\\
   & \qquad 0\le\phi_n< 2\pi,\qquad\quad \quad \forall n \in \mathcal{N}. \label{PS1_C3}
\end{align}
\end{subequations}
\end{minipage}
\noindent\makebox[\textwidth]{\rule{\textwidth}{0.4pt}}
\vspace{-0.5in}
\end{figure*}
\subsection{URLLC RIS phase-shift matrix $\boldsymbol{\Phi}_{\rm u}^*$}
\label{urllc_alone} 
As discussed above, the goal of configuring the RIS is to add a degree of freedom for the BS to improve the URLLC reliability during each mini-time slot. Accordingly, this approach consists of exploiting the CSI of the coexisting URLLC users in designing the RIS phase-shift matrix. As such, the RIS phase-shift matrix $\boldsymbol{\Phi}_{\rm u}^*$ is designed at the beginning of the time slot with the aim of enhancing the channel gains of all URLLC users in the network. Therefore, the problem of designing the RIS phase-shift matrix $\boldsymbol{\Phi}_{\rm u}^*$ has the objective of maximizing the minimum URLLC channel gain, which can be formulated as shown in problem $P_{3}$ in \eqref{P3} on top of this page. The solution of $\mathcal{P}_3$ is detailed in Appendix B. 
 \subsection{Joint URLLC-eMBB RIS phase-shift matrix $\boldsymbol{\Phi}_{\rm e,u}^*$} The phase-shift matrix $\boldsymbol{\Phi}_{\rm u}^*$ obtained by solving problem $\mathcal{P}_{3}$ enhances the performance of the URLLC traffic. However, this may highly impact the performance of the eMBB traffic. Alternatively, the use of a unified RIS phase-shift matrix that can improve the performance of the URLLC load while reducing the degradation on the performance of the eMBB traffic is highly desired. Similar to problem $\mathcal{P}_3$, an RIS configuration, represented by the phase-shift matrix configuration $\boldsymbol{\Phi}_{\rm e,u}^*$ is designed at the beginning of the time slot with the aim of enhancing the channel gains of all coexisting eMBB and URLLC users in the network. Accordingly, the problem is formulated as maximizing the minimum channel gain of all coexisting eMBB and URLLC users, which is formulated as in problem $P_{4}$ in \eqref{P4} on top of this page, where $t_{{\rm BS}, x} = h_{{\rm BS}, x}$, if $x \in \mathcal{E}_f$ and $t_{{\rm BS}, x} = g_{{\rm BS}, x-E_f}$ if $x\in \left\{ E_f+1,\dots, E_f+U\right\}$, 
and $\mathbf{t}_{{\rm RIS}, x} = \mathbf{h}_{{\rm RIS}, x}$, if $x \in \mathcal{E}_f$ and $\mathbf{t}_{{\rm RIS}, x} = \mathbf{g}_{{\rm RIS}, x-E_f}$ if $x\in \left\{ E_f+1,\dots, E_f+U\right\}$. The solution of problem $\mathcal{P}_4$ is detailed in Appendix B.
\begin{figure*}
\begin{minipage}{.5\textwidth}
\begin{subequations}
\label{P3}
\begin{align}
  \mathcal{P}_3:  &\max_{\boldsymbol{\Phi}_{\rm u} }\min_{1 \leq u \leq U} |g_{{\rm BS}, u}+  \mathbf{g}_{{\rm RIS},u}^H \boldsymbol{\Phi}_{\rm u} \,\mathbf{f}_{{\rm  BS},{\rm  RIS}}|^2\\
 &\text{s.t.}\quad 0\le\phi_n< 2\pi,\quad \forall n \in \mathcal{N}.
\end{align}
\end{subequations}
\end{minipage}
\begin{minipage}{.5\textwidth}
\begin{subequations}
\label{P4}
\begin{align}
  \mathcal{P}_4:  &\max_{\boldsymbol{\Phi}_{\rm e,u} }\min_{1 \leq x \leq E_f+U} |t_{{\rm BS}, x}+  \mathbf{t}_{{\rm RIS},x}^H \boldsymbol{\Phi}_{\rm e,u} \,\mathbf{f}_{{\rm  BS},{\rm  RIS}}|^2\\
 &\text{s.t} \quad 0\le\phi_n< 2\pi,\quad \forall n \in \mathcal{N}.
\end{align}
\end{subequations}
\end{minipage}
\noindent\makebox[\textwidth]{\rule{\textwidth}{0.4pt}}
\end{figure*}
\subsection{URLLC Allocation}
\indent The optimization problems $\mathcal{P}_1$, $\mathcal{P}_3$ and $\mathcal{P}_4$ will be solved in parallel at the beginning of each time slot. Once the three RIS configurations $\left\{ \boldsymbol{\Phi}_{\rm e}^*,\boldsymbol{\Phi}_{\rm u}^*,\boldsymbol{\Phi}_{\rm e,u}^*\right\}$ are obtained at the beginning of each time-slot, they will be communicated to the RIS controller. Afterwards, within each mini-time slot, a joint power/resource allocation problems for URLLC packets is solved for each candidate RIS phase-shift matrix in the set $\left\{ \boldsymbol{\Phi}_{\rm e}^*,\boldsymbol{\Phi}_{\rm u}^*,\boldsymbol{\Phi}_{\rm e,u}^*\right\}$. Specifically, for all $m \in \mathcal{M}$, the URLLC allocation problem $\mathcal{P}_2^m$ will be solved when the RIS phase-shift matrix $\boldsymbol{\Phi}^m$ is one of the pre-computed RIS phase-shift matrices, i.e., $\boldsymbol{\Phi}^m \in \left\{ \boldsymbol{\Phi}_{\rm e}^*, \boldsymbol{\Phi}_{\rm u}^*, \boldsymbol{\Phi}_{\rm e,u}^*\right\}$. For each of these three cases, the URLLC allocation problem is reduced to a simple joint power/frequency allocation problem and the resulting three optimization problems can solved in a parallel. After doing so, the BS selects the best RIS configuration $ \boldsymbol{\Phi}^{m^*} \in \left\{ \boldsymbol{\Phi}_{\rm e}^*, \boldsymbol{\Phi}_{\rm u}^*, \boldsymbol{\Phi}_{\rm e,u}^*\right\}$, along with the corresponding optimal power and frequency allocation policies for the URLLC traffic, such that the admitted URLLC packets is maximized. Afterwards, at each mini-time slot $m$, for all $m \in \mathcal{M}$, the BS sends a control signal to the RIS in order to switch the RIS configuration to the best obtained configuration $\boldsymbol{\Phi}^{m^*}$. The overall URLLC allocation procedure in each mini-time slot is presented in \textbf{Algorithm 1}. The remaining now is how to obtain the optimal frequency and power allocation policies for the URLLC traffic at each mini-time slot when the RIS phase-shift matrix is fixed, which is detailed in the following section.
\begin{algorithm}[t!]
\SetAlgoLined
 \For{$\boldsymbol{\Phi}^m\in \left\{ \boldsymbol{\Phi}_{\rm e}^*, \boldsymbol{\Phi}_{\rm u}^*, \boldsymbol{\Phi}_{\rm e,u}^*\right\}$}{
 Solve problem $\mathcal{P}_{2}^m$ and get the optimal $ {\mathbf{p}_{\rm L}^m}^*(\boldsymbol{\Phi}^m),{\mathbf{k}^m}^*(\boldsymbol{\Phi}^m) , {\mathbf{I}^m}^*(\boldsymbol{\Phi}^m) $\;
 }
  - ${\boldsymbol{\Phi}^m}^*=\arg \max_{\boldsymbol{\Phi}^m}{\Big \{}{\mathbf{k}^m}^* (\boldsymbol{\Phi}^m)|\,\boldsymbol{\Phi}^m\in \left\{ \boldsymbol{\Phi}_{\rm e}^*, \boldsymbol{\Phi}_{\rm u}^*, \boldsymbol{\Phi}_{\rm e,u}^*\right\}{\Big \}}$\;
  - $ {\mathbf{p}_{\rm L}^m}^*={\mathbf{p}_{\rm L}^m}^*({\boldsymbol{\Phi}^m}^*) $, 
    $ {\mathbf{k}^m}^*={\mathbf{k}^m}^*({\boldsymbol{\Phi}^m}^*) $,
   and $ {\mathbf{I}^m}^*={\mathbf{I}^m}^*({\boldsymbol{\Phi}^m}^*) $\;
    \caption{Proposed Algorithm  }
  \label{alg:1}
\end{algorithm}
\setlength{\textfloatsep}{4pt}
\section{The URLLC allocation problem for fixed RIS phase-shift configuration}
\label{urllc_sec}
\subsection{URLLC Resource Allocation}
In this section, the RIS phase-shift matrix $\boldsymbol{\Phi}^m$, for all $m \in \mathcal{M}$, is fixed, i.e.,  $\boldsymbol{\Phi}^m\in \left\{ \boldsymbol{\Phi}_{\rm e}^*, \boldsymbol{\Phi}_{\rm u}^*, \boldsymbol{\Phi}_{\rm e,u}^*\right\}$. In this case, for all $m \in \mathcal{M}$, the URLLC allocation problem $\mathcal{P}_2^m$ is reduced to a joint power/frequency allocation problem. To simplify problem $\mathcal{P}_2^m$, we consider disjoint optimization integer vectors $\mathbf{I}_{E_f}^m = \left[I_1^m,I_2^m,\dots,I_{E_f}^m \right]$ and $\mathbf{I}_L^m = \left[I_1^m,I_2^m,\dots,I_{L^m}^m \right]$ for the punctured eMBB RBs and the RBs allocated to the URLLC packets, respectively, i.e, using the change of variables $I_l^m=\sum_{e=1}^{E_f} I_{e,l}^m$ and $I_e^m=\sum_{l=1}^{L^m} I_{e,l}^m$, for all $e \in \mathcal{E}_f$ and $l \in \mathcal{L}^m$. Then, problem $\mathcal{P}_{2}^m$ can be reformulated as
\textcolor{black}{
\begin{subequations}
\begin{align}
  \mathcal{P}_{2,1}^m:  \quad &\max_{\mathbf{p}_{\rm L}^m, \mathbf{I}_{E_f}^m,\mathbf{I}_L^m ,\mathbf{k}^m } \,\,[f_1(\mathbf{k}^m),\,-f_2(\mathbf{I}_{E_f}^m)], \label{P2,1_obj} \\
    &\text{s.t.}  \quad \eqref{P2_c1}-\eqref{P2_c5},\\
    & \qquad I_{e}^m\in \{0,1,\dots,b\},\,\, \forall e\in \mathcal{E}_f, \label{P2,1_c1}\\
     & \qquad I_{l}^m\in \{0,1,\dots,B\},\,\, \forall l\in \mathcal{L}^m, \label{P2,1_c2}\\
    &\qquad \sum_{l=1}^{L^m} I_{l}^m= \sum_{e=1}^{E_f} I_{e}^m \label{P2,1_c3}.
\end{align}
\end{subequations}}
 Constraint (\ref{P2,1_c3}) guarantees that the number of punctured eMBB resources is equal to the number the resources allocated to the URLLC traffic. Problem $\mathcal{P}_{2,1}^m$ is a mixed integer non-linear problem (MINLP), and it is a non-convex problem due to the non-convex constraints \eqref{P2_c1} and \eqref{P2_c2}. Using the fact that the power allocated to the eMBB users are already optimized at the beginning of the time slot, then $\sum_{e=1}^{E_f} b\, p_e=P_{\rm BS}$. Hence, for all $m \in \mathcal{M}$ and $e \in \mathcal{E}_f$, by manipulating constraints \eqref{P2_c2} and \eqref{P2,1_c1}, one can reach that the maximum available RBs for allocating URLLC load over eMBB user $e$ at mini-time slot $m$ is
\begin{equation}
    I_e^{{\rm max},m}=\min\left({\Bigg\lfloor} b\,\left(1-\frac{r_{e,\rm th}^{'}}{r_e\left(\boldsymbol{\Phi}^m,p_e\right)}\right){\Bigg\rfloor},b\right),
\end{equation}
where for all $x\in \mathcal{R}$, $\lfloor x \rfloor$ represents the greatest integer less than or equal $x$. From constraint (\ref{P2_c1}), one can see that, the optimal power allocation for the $l$th URLLC packet given a fixed number of allocated RBs $I_l$, for all $l \in \mathcal{L}^m$ and $m \in \mathcal{M}$, is given by
\begin{equation}
\label{optimal_p}
  {p_l^m}^*(I_l^m)=\frac{ \left( e^{\frac{k_l^m \, c_{\rm th}}{\log(2)I_l^m}}-1\right)}{\alpha_l^m} ,
\end{equation}
where $\alpha_l^m=\frac{|g_{{\rm BS}, l}+  \mathbf{g}_{{ \rm RIS},l}^H \boldsymbol{\Phi}^m \,\mathbf{f}_{{\rm  BS},{\rm  RIS}}|^2}{\,\sigma^2\Gamma_{\rm URLLC}}$. Based on this, for all $m \in \mathcal{M}$, problem $\mathcal{P}_{2,1}^m $ can be equivalently transformed to
\begin{subequations}
\begin{align}
  \mathcal{P}_{2,2}^m:  \quad &\max_{\mathbf{I}_{E_f}^m,\mathbf{I}_L^m ,\mathbf{k}^m } \,\,[f_1(\mathbf{k}^m),\,-f_2(\mathbf{I}_{E_f}^m)], \label{P2,2_obj} \\
    &\text{s.t.}  \quad \eqref{P2_c4}-\eqref{P2_c5},\eqref{P2,1_c1}-\eqref{P2,1_c3}\\
    &\qquad\sum_{l=1}^{L^m} I_l^m {p_l^m}^*(I_l^m)\le \sum_{e=1}^{E_f} I_e^m p_e,\,\,\qquad\forall l\in\mathcal{L}^m. \label{P2,2_c1}
\end{align}
\end{subequations}
\textcolor{black}{The proposed solutions of $\mathcal{P}_{2,2}^m$ are provided in the following subsection. Specifically, we proposed two algorithms for the URLLC allocation problem: an optimization-based algorithm and a Heuristic algorithm. The Heuristic algorithm aims to overcome the computational complexity of the optimization-based algorithm such the URLLC latency requirements are met.}  
\subsection{Proposed Algorithms}
\subsubsection{Optimization-based URLLC allocation}
\begin{algorithm}[t!]
\SetAlgoLined
 - \textbf{Solve}  problem $\mathcal{P}_{2,4}^m $and get the optimal ${\mathbf{k}^m}^*$ and the corresponding ${\hat{\mathcal{L}}}^m\subset{\mathcal{L}}^m $ \;
 - \textbf{Solve} problem $\mathcal{P}_{2,5}^m $ for and get the optimal ${{\mathbf{I}_L^m}^*}$, ${{\mathbf{I}_{E_f}^m}^*}$, \;
- \textbf{Evaluate} the optimal ${{\mathbf{p}_L^m}^*}$ from  \eqref{optimal_p}\;
- \textbf{Evaluate}  ${\mathbf{I}^m}^* $ by distributing ${\mathbf{I}_L^m}^*$ over ${{\mathbf{I}_{E_f}^m}^*}$\;
 \caption{Optimization-based URLLC allocation algorithm  }
  \label{alg:OTOPP}
\end{algorithm}
\setlength{\textfloatsep}{4pt}
\textcolor{black}{For all $m \in \mathcal{M}$, problem $\mathcal{P}_{2,2}^m$ is a multi objective problem which aims at concurrently maximizing the number of admitted URLLC packets and minimizing the eMBB rate loss at mini-time slot $m$. Due to its latency and reliability constraints, it is important to mention here that the objective of maximizing the number of admitted URLLC packets has a higher priority than the one of minimizing the eMBB loss. Hence, an efficient solution is hard to be obtained by solving problem $\mathcal{P}_{2,2}^m$ directly. Alternatively, for all $m \in \mathcal{M}$, we decompose problem $\mathcal{P}_{2,2}^m$ into two sub-problems, namely, a URLLC admission problem and a URLLC allocation problem.} The first sub-problem aims at maximizing the admission of URLLC packets ${\sum_{l=1}^{L^m} k_l^m}$, and it is expressed as shown in problem $\mathcal{P}_{2,3}^m$ in \eqref{P23} on top of this page.
\begin{figure*}
\begin{minipage}{.5\textwidth}
\begin{subequations}
\label{P23}
\begin{align}
  \mathcal{P}_{2,3}^m:  \quad &\max_{\mathbf{I}_{E_f}^m,\mathbf{I}_L^m ,\mathbf{k}^m } \,\,f_1(\mathbf{k}^m)\label{P2,3_obj} \\
    &\text{s.t.}  \quad \eqref{P2_c4}-\eqref{P2_c5},\eqref{P2,1_c1}-\eqref{P2,1_c3},\eqref{P2,2_c1}.
\end{align}
\end{subequations}
\end{minipage}
\begin{minipage}{.5\textwidth}
\begin{subequations}
\label{P24}
\begin{align}
  \mathcal{P}_{2,4}^m:  \quad &\min_{\mathbf{I}_{E_f}^m,\mathbf{I}_L^m} \,\,f_2(\mathbf{I}_{E_f}^m)\label{P2,4_obj} \\
    &\text{s.t.}  \quad \eqref{P2,1_c1}-\eqref{P2,1_c3},\eqref{P2,2_c1}.
\end{align}
\end{subequations}
\end{minipage}
\noindent\makebox[\textwidth]{\rule{\textwidth}{0.4pt}}
\end{figure*}
Once the optimal number of admitted URLLC packets ${\mathbf{k}^m}^*$ is obtained at each mini-time slot $m$, for all $m \in \mathcal{M}$, then the set of the URLLC packets $\hat{\mathcal{L}}^m\subseteq {\mathcal{L}}^m$ that can be allocated is defined, i.e., $\hat{\mathcal{L}}^m = \left\{l \in \mathcal{L}^m| {k}^m_l=1\right\}$. Then, the optimal frequency resources for admitting the URLLC packets, i.e., the optimal frequency resources to be punctured from the eMBB users, are determined based on the chosen allocation strategy $\pi$. The associated optimization problem is formulated as shown in problem $\mathcal{P}_{2,4}^m$ in \eqref{P23} on top of this page. Although problems $\mathcal{P}_{2,3}^m$ and $\mathcal{P}_{2,4}^m$ can be solved optimally using an integer optimization solver \cite{mosek}, the computational time is very high. Hence, we can solve these problems by relaxing the integer variables $\mathbf{I}_L^m$ and $\mathbf{I}_{E_f}^m$ to be continuous. Then, the resulting solutions are rounded to get the optimal integer values ${\mathbf{I}_L^m}^*$ and ${\mathbf{I}_{E_f}^m}^*$. Hence, the optimal ${{\mathbf{p}_L^m}^*}$ is evaluated from \eqref{optimal_p}. Based on the above, the solution approach of problem $\mathcal{P}_{2,1}^m$ is summarized in \textbf{Algorithm \ref{alg:OTOPP}}. Finally, by distributing the URLLC packets over the punctured eMBB RBs ${\mathbf{I}_{E_f}^m}^*$ based on their required RBs ${\mathbf{I}_L^m}^*$, the solution of the original integer problem $\mathcal{P}_{2}^m$ can be obtained.
\subsubsection{Heuristic URLLC allocation algorithm}
{
\begin{algorithm}[t!]
\setstretch{0.9}
\SetAlgoLined
Initiate $\mathbf{I}^m$ and $\mathbf{p}_L^m$\;
Evaluate $\mathbf{I}^{{\rm max},m}$ and $\boldsymbol{\beta}^{m,\pi} $\;
\textbf{Sort} URLLC users based on the channel gain in descending order\;
\textbf{Sort} eMBB users based on $\boldsymbol{\beta}^{m,\pi} $in ascending order \;
 \For{ $l=1\rightarrow L^m$}
 {
$c_l^{temp}=0$\; 
Boolean=0,\, URLLC binary variable variable\;   
     \For{$e=1\rightarrow E_f$}
       { 
         $I_{e,l}^{temp}=\lceil\frac{c_{th}-c_l^{temp}}{c_l(p_e)}\rceil $\;
       \eIf {$I_{e,l}^{temp}\le I_{e}^{{\rm max},m}$ }
       {  $I_{e,l}^m= I_{e,l}^{temp}$\;
       Boolean=1;
       break\;
       }{$I_{e,l}^m=I_{e}^{{\rm max},m}$\;
       $c_l^{temp}=c_l^{temp}+I_{e,l}^m*c_l(p_e)$\;
       }
    }
    \eIf{boolean=1}
    {$p_l=\frac{\sum_{e=1}^{E_f} I_{e,l} p_e}{\sum_{e=1}^{E_f} I_{e,l}}$\;
    update $\mathbf{I}^{{\rm max},m}$\;}{break\;}
}
 \caption{ Heuristic URLLC allocation algorithm }
  \label{alg:2}
\end{algorithm}}
\setlength{\textfloatsep}{6pt}
The URLLC traffic requires strict latency requirements which is less than $1~$ msec, and hence, a low complexity algorithm is essential to allocate the incoming URLLC packets. Accordingly, in this part, we develop a low complexity Heuristic algorithm that exploits two observations. First, because the channel gains of the URLLC users are known for fixed RIS phase-shift matrix $\boldsymbol{\Phi}^m$, the URLLC packets with good channel conditions need less frequency and power resources than those with bad channel conditions. Accordingly, allocating the URLLC packets with good channel conditions before those with bad channel conditions will enhance the URLLC admission rate. Second, as was shown in section \ref{problem formulation}, for all $m\in \mathcal{M}$, the eMBB users with low allocation weights have a higher chance to be punctured than those with high allocation weights. Accordingly, the proposed approach starts by sorting the URLLC packets in descending order based on their channel conditions. Specifically, at each mini-time slot $m$, for all $m \in \mathcal{M}$, and when the RIS phase-shift matrix $\boldsymbol{\Phi}^m$ is fixed, let  $g_l^m=|g_{{\rm BS}, l}+  \mathbf{g}_{{\rm RIS},l}^H \boldsymbol{\Phi}^m \,\mathbf{f}_{{\rm  BS},{\rm  RIS}}|^2$ be the channel gain of the URLLC user associated to the $l$th URLLC packet, for $l\in \mathcal{L}^m$. Then, the sorted channel gain of the URLLC users can be expressed as $\textbf{g}^m =\{g_{(1)}^m,g_{(2)}^m,\dots,g_{(L^m)}^m\}$ where, for all $l \in \mathcal{L}^m$, $g_{(l)}^m$ is the $l$th highest channel gain. Similarly, the eMBB users are sorted in a ascending order based on their allocation weights of the URLLC load allocation as $\hat{\boldsymbol{\beta}}^{m,\pi} =\{{\beta}^{m,\pi}_{(1)},{\beta}^{m,\pi}_{(2)},\dots,{\beta}^{m,\pi}_{(E_f)}\}$ where, for all $e \in \mathcal{E}_f$, ${\beta}^{m,\pi}_{(l)}$ is the $l$th lowest allocation weight. Afterwards, for all $l \in \mathcal{L}^m$, the algorithm assumes zero data rate $c_l$ for the $l$th URLLC packet. Then, the proposed approach allocates resources to the $l$th URLLC packet by iterating over the ordered eMBB users and considering that the power allocated to $l$th URLLC packet on the punctured RB is equal to that of the eMBB user within each iteration. The algorithm continues the iterative procedure until the rate requirement of the $l$th URLLC packet is satisfied at a given eMBB user. Precisely, for all $l \in \mathcal{L}^m$, the algorithm iterates over the eMBB users and checks if the cumulative data rate of the $l$th URLLC packet at each eMBB user is higher than the rate threshold $c_{\rm th}$, i.e.,
\begin{equation}\label{eq:condition}
\sum_{i=1}^e I_{i,l}^m c_l(p_i)= \sum_{i=1}^{e} I_{i,l}^m\times \log(1+g_l\,{p_i I_{i,l}^m})\ge c_{\rm th},
\end{equation}
where $e$ is the index of the eMBB user under checking. Once the frequency resources $\left(I_{i,l}^m\right)_{1\leq i \leq e}$ are allocated to $l$the URLLC packet, the power that will be allocated to the $l$th URLLC packet per RB is given by $p_l^m=\frac{\sum_{i=1}^{e} I_{i,l}^m\, p_e}{\sum_{i=1}^{e}I_{i,l}^m}$. This is basically due to the facts that the condition in \eqref{eq:condition} guarantees that the URLLC rate requirement is satisfied and that the rate function is concave, i.e.,
\begin{equation}\label{6mini_slot_rate}
    \log\left(1+g_lp_l^m\right) \sum_{i=1}^{e} I_{i,l} = \log\left(1+g_l\,\frac{\sum_{i=1}^{e} p_i I_{i,l}^m}{ \sum_{i=1}^{e} I_{i,l}^m}\right) \sum_{i=1}^{e} I_{i,l} \sum_{i=1}^{e} I_{i,l}^m\times \log(1+g_l\,{p_i I_{i,l}}) \ge c_{\rm th}.
\end{equation} 
These steps are repeated for all URLLC packets while satisfying the eMBB QoS. Otherwise, the URLLC packet is dropped. \textcolor{black}{ Finally, \textbf{Algorithm 3} presents the details steps of the proposed approach, where it consists of two loops, an inner loop and an outer loop of $L^m$ and $E_f$ iterations, respectively. This algorithm has a polynomial time complexity with respect to the number of URLLC packet $L^m$ and the worst-case time complexity of the algorithm is $O(L^m \times E_f)$.}
\section{Simulation results}\label{Simulation results}
\subsection{Simulation Setup}  
\indent In this section, we perform various simulations to evaluate the performance of the proposed scheme. In this simulation environment, the wireless network consists of one BS and one RIS. The coverage area of the BS is assumed to be 110 meters. The BS serves $E=8$ eMBB users and $U$ (in the range $[5,80]$) URLLC users are distributed uniformly at random over the coverage area of the BS. In order to increase the BS coverage, the RIS is located 20 meters away from it. The number of RIS reflecting elements $N$ is a key parameter in the performance of the proposed scheme. Hence, we vary $N$ in the range $[ 10,50]$. Each time slot consists of $M = 7$ mini-time slots. We consider both the large-scale fading and the small-scale fading for all communication links. Particularly, the large scale fading for the direct and the cascaded links are modeled as $P_0=\alpha_0(d_{\rm BS,e/u})^{-\varrho_{0}}$ and $P_1=\alpha_{1}(d_{\rm BS,RIS})^{-\varrho_{1}}(d_{\rm BS,e/u})^{-\varrho_{2}}$, respectively, where $d_{\rm BS,e/u}$, $d_{\rm BS,RIS}$ and $d_{\rm BS,RIS}$ are the distances of BS-users, BS-RIS, and RIS-users links, respectively. $\varrho_{0} = 3.5$, $\varrho_{1} = 2.2$ and $\varrho_{2} = 2.8$ are the
the path loss exponents of BS-users, BS-RIS, and RIS-users links, respectively. Also, $\alpha_0 = -30$ dB and $\alpha_1 = -40$ dB are the path loss at the reference distance for the direct links and the cascaded links, respectively. On the other hand, the small-scale fading for all channels is modeled as $f=\sqrt{\frac{\kappa}{1+\kappa}}\, f^{\text {LoS }}+\sqrt{\frac{1}{1+\kappa}}\, f^{\text {NLoS }}$, where $\kappa$ is the Rician factor, $f^{\text {LoS }}$ is the line-of-sight component and $f^{\text {NLoS }}$ is non-line-of-sight component, which follows a Rayleigh distribution with a scale parameter equals to one  \cite{zuo2020reconfigurable}. The communication links between the BS and the cellular users and between the RIS and the cellular users are assumed to have a quasi-static flat-fading Rayleigh channel. The links between the BS and the RIS elements are assumed to have a Rician channel model with a Rician factor $\kappa=10$ \cite{zuo2020reconfigurable}. The remaining system parameters are summarized in Table \ref{Table:Simulation} \cite{9174801,zuo2020reconfigurable,zhang2020joint,ghanem2020joint}. The simulation results are performed over $2*10^3$ independent Monte-Carlo realizations on the channel gains of all cellular users.\\
\begin{table}[!t]
\caption{\label{Table:Simulation} Simulation parameters}
\renewcommand{\arraystretch}{.4} 
\setlength{\tabcolsep}{0.05cm} 
\centering 
\begin{tabular}{|c|c|c|c|c|c|}
 \hline
 Parameter&Symbol& Value &Parameter&Symbol& Value  \\
 \hline
  Power budget at the BS&$P_{\rm BS}$& 33~dBm &  Noise power& $\sigma^2$& $-97.5$ dBm\\
 \hline
 URLLC packet arrival rate  &$\lambda_u$ & 0.7 packet/msec &   Mini-slot duration&$\tau$ & 0.143 ms  \\
 \hline
   Number of resource blocks&$B$& 96 &   Resource block bandwidth & W& 180 kHz \\
 \hline
     eMBB block error probability & $\epsilon_{\rm eMBB}$& $10^{-1}$  & eMBB rate threshold &$r_{\rm th}$& $1~$Mb/sec\\ 
      \hline
    URLLC block error probability & $\epsilon_{\rm URLLC}$& $10^{-6}$ & URLLC packet size&$\zeta$ & 256 bits \\
 \hline
\end{tabular}
\vspace{-.27 in}
\end{table}
\indent In order to show the performance of the different possible RIS configurations on the performance of both the URLLC and the eMBB traffic, we introduce the following four schemes:
 \begin{itemize}
     \item Proposed Scheme-1: The RIS phase-shift matrix $\boldsymbol{\Phi}_{\rm e}^*$ is used to serve the eMBB and the URLLC during the entire time slot, i.e., at each mini-time slot.
     \item Proposed Scheme-2: The RIS phase-shift matrix $\boldsymbol{\Phi}_{\rm u}^*$ is used whenever a URLLC packet has arrived. Accordingly, at each mini-time slot, if a URLLC packet exists, the RIS configuration $\boldsymbol{\Phi}_{\rm u}^*$ is used. Otherwise, the configuration $\boldsymbol{\Phi}_{\rm e}^*$ is used.
     \item Proposed Scheme-3: The RIS phase-shift matrix $\boldsymbol{\Phi}_{\rm e, u}^*$ is used whenever a URLLC packet has arrived. Accordingly, at each mini-time slot, if a URLLC packet exists, the RIS configuration $\boldsymbol{\Phi}_{\rm e, u}^*$ is used. Otherwise, the configuration $\boldsymbol{\Phi}_{\rm e}^*$ is used.
     \item {Selected RIS configuration}: In each mini-time slot, the BS computes the maximum URLLC admitted packets and the lowest eMBB rate loss for the RIS configurations $\boldsymbol{\Phi}_{\rm e}^*$, $\boldsymbol{\Phi}_{\rm u}^*$ and $\boldsymbol{\Phi}_{\rm e,u}^*$. Then, as shown in \textbf{Algorithm 1}, the BS will select the phase-shift matrix that maximizes the number of admitted URLLC packets.
 \end{itemize}
 \vspace{-0.2in}
 \subsection{Performance Evaluation of the URLLC Allocation Strategies}
\begin{figure}[t]
\centering
\begin{subfigure}[b]{.32\columnwidth}
\centering
\includegraphics[width=1\columnwidth,draft=false]{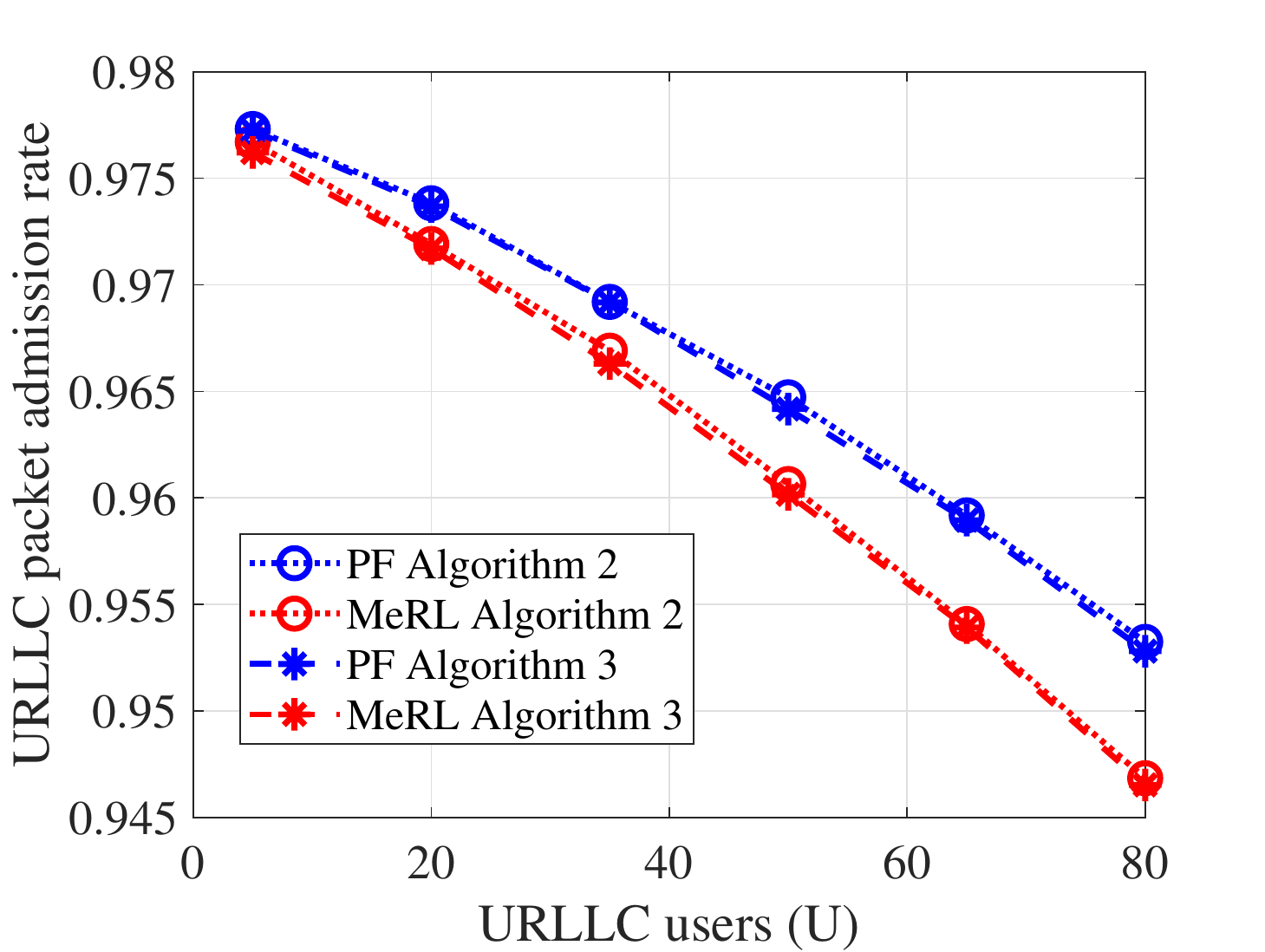}
\caption{URLLC packets admission rate against the number URLLC users. }
\label{Fig:opt_reliabiy}
\end{subfigure}~
\begin{subfigure}[b]{.3\columnwidth}
\centering
\includegraphics[width=1\columnwidth,draft=false]{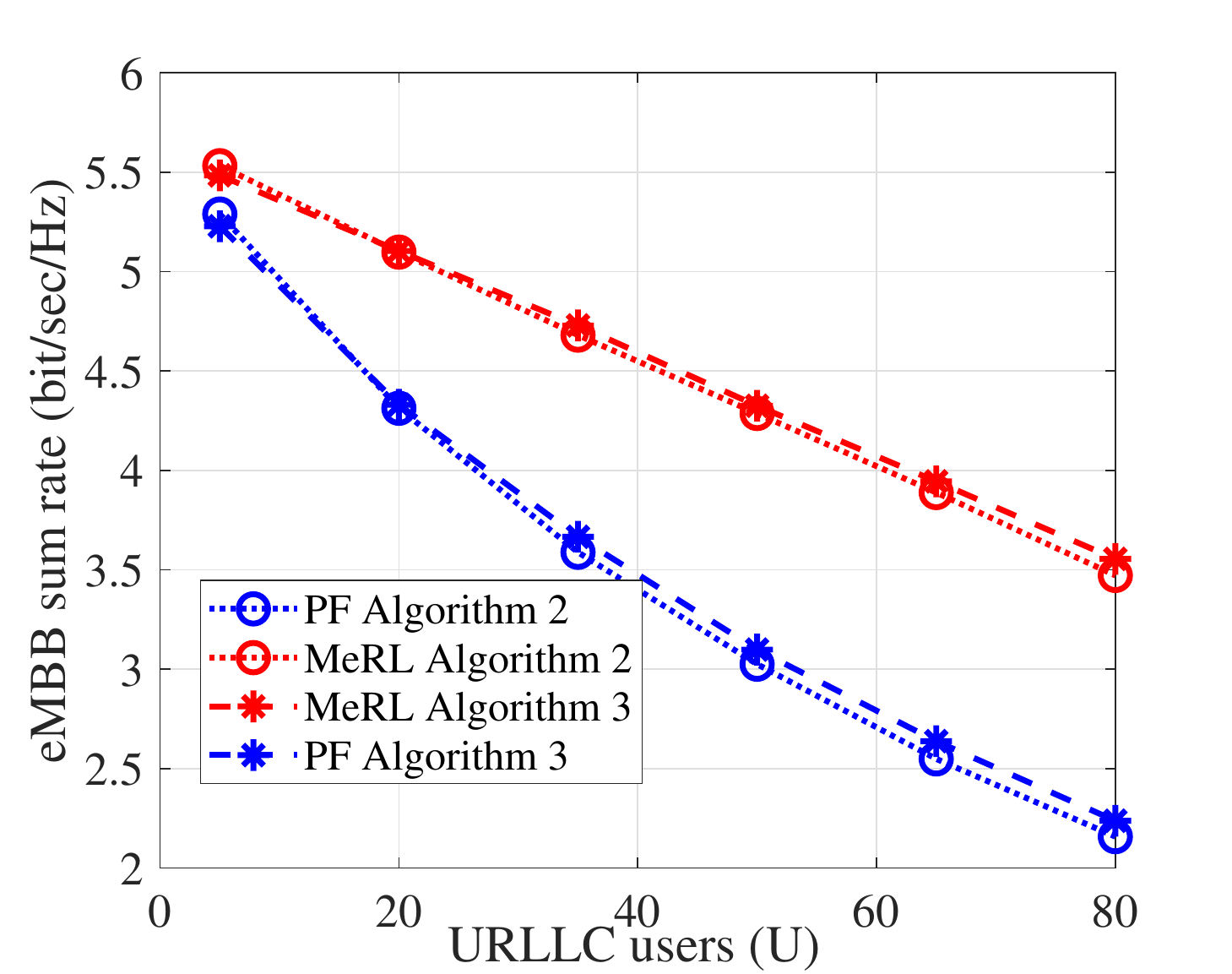}
\caption{eMBB sum rate against the number URLLC users.}
\label{Fig:opt_RATE}
\end{subfigure}
~
\begin{subfigure}[b]{.32\columnwidth}
\centering
\includegraphics[width=1\columnwidth,draft=false]{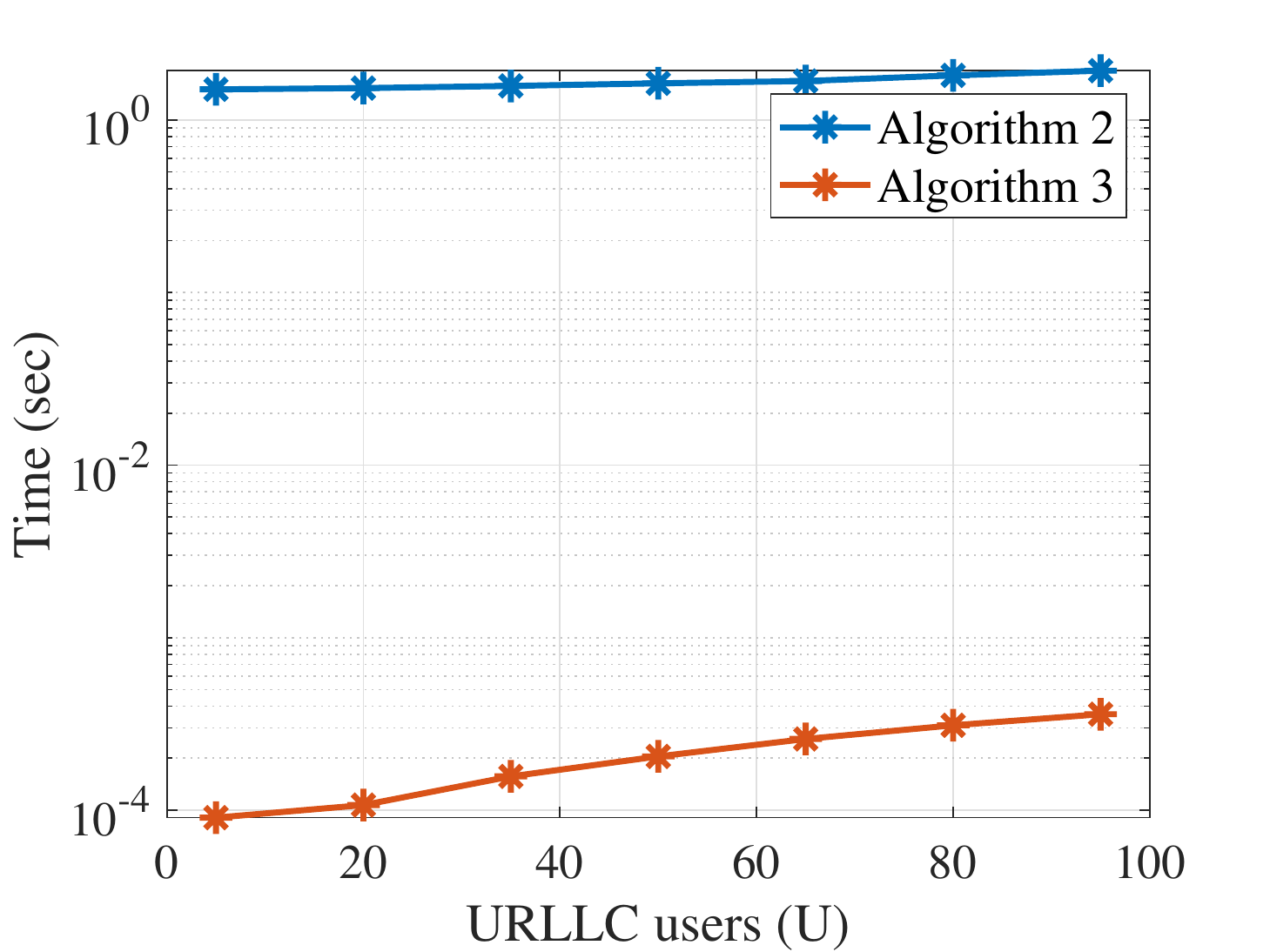}
\caption{Algorithm time complexity vs number of URLLC users.}
\label{Fig:complexity}
\end{subfigure}
\caption{Performance comparison between \textbf{Algorithm 2} and \textbf{Algorithm 3}.}
\vspace{-.1 in}
\end{figure}
Fig.\ref{Fig:opt_reliabiy} and Fig.\ref{Fig:opt_RATE} illustrate the performance of \textbf{Algorithm 2} and \textbf{Algorithm 3} in terms of both the URLLC packets admission rates and the eMBB sum rate, respectively, while varying the number of URLLC users (URLLC load). {We observe that as the URLLC load increases, the URLLC admission rate starts to reduce, which is due to the following. First, the BS has limited power and frequency resources and is required to protect the QoS of the already admitted eMBB; therefore, at each mini slot, it decides which resources can be punctured to admit the incoming URLLC packets. As more packets arrive, the available resources which can be punctured without impacting the eMBB service may not suffice to admit all the URLLC load and therefore some packets get rejected; indeed, both algorithms (and policies) exhibits similar performance decay as observed from the figures.}  
 Furthermore, it can be seen from Fig. \ref{Fig:opt_reliabiy} that the PF URLLC allocation strategy achieves better URLLC packets admission rate than the MeRL allocation strategy. The reason being that by puncturing the eMBB users with low data rates, the eMBB resources available for puncturing become limited to accommodate the high URLLC load. Furthermore, Fig. \ref{Fig:opt_RATE} shows that the MeRL URLLC allocation strategy achieves better eMBB sum rate than that of the PF URLLC allocation. The reason behind that is the punctured resources resulting from the MeRL approach are belonging to the eMBB users with low data rates rather than those with high data rates. Since, the PF achieves better URLLC packets admission rate than the MeRL, we consider the PF for the rest of simulations. { On the other hand, Fig. \ref{Fig:opt_reliabiy} and Fig. \ref{Fig:opt_RATE} show that \textbf{Algorithm 2} and \textbf{Algorithm 3} have broadly the same performance in terms of both the URLLC packets admission rates and the eMBB sum rate, respectively.} However, Fig. \ref{Fig:complexity} illustrates the run time complexity of \textbf{Algorithm 2} and \textbf{Algorithm 3}. It can be shown that \textbf{Algorithm 2} computation time is in the order of one second which can violate the URLLC latency requirements. However, \textbf{Algorithm 3} has a lower compute time which is around one mini-time slot, $0.143$ millisecond. This low operating time makes \textbf{Algorithm 3} favorable in practice. Moreover, the processing time of the \textbf{Algorithm 3} can be further improved when using more computing resources available at the network edge.\footnote{The algorithm was implemented in Matlab using a machine with the following characteristics: System Type: x64-based PC Processor: Intel(R) i7-4510U CPU @2GHz, 8 Gigabyte RAM.} Since, \textbf{Algorithm 3} has very low time complexity and almost similar performance to \textbf{Algorithm 2}, we consider \textbf{Algorithm 3} for the rest of our simulations.
 \vspace{-0.2in}
 \subsection{Impact of the Number of RIS Elements}
 \begin{figure}[t]
\centering
\begin{subfigure}[b]{.35\columnwidth}
\centering
\includegraphics[width=1\columnwidth,draft=false]{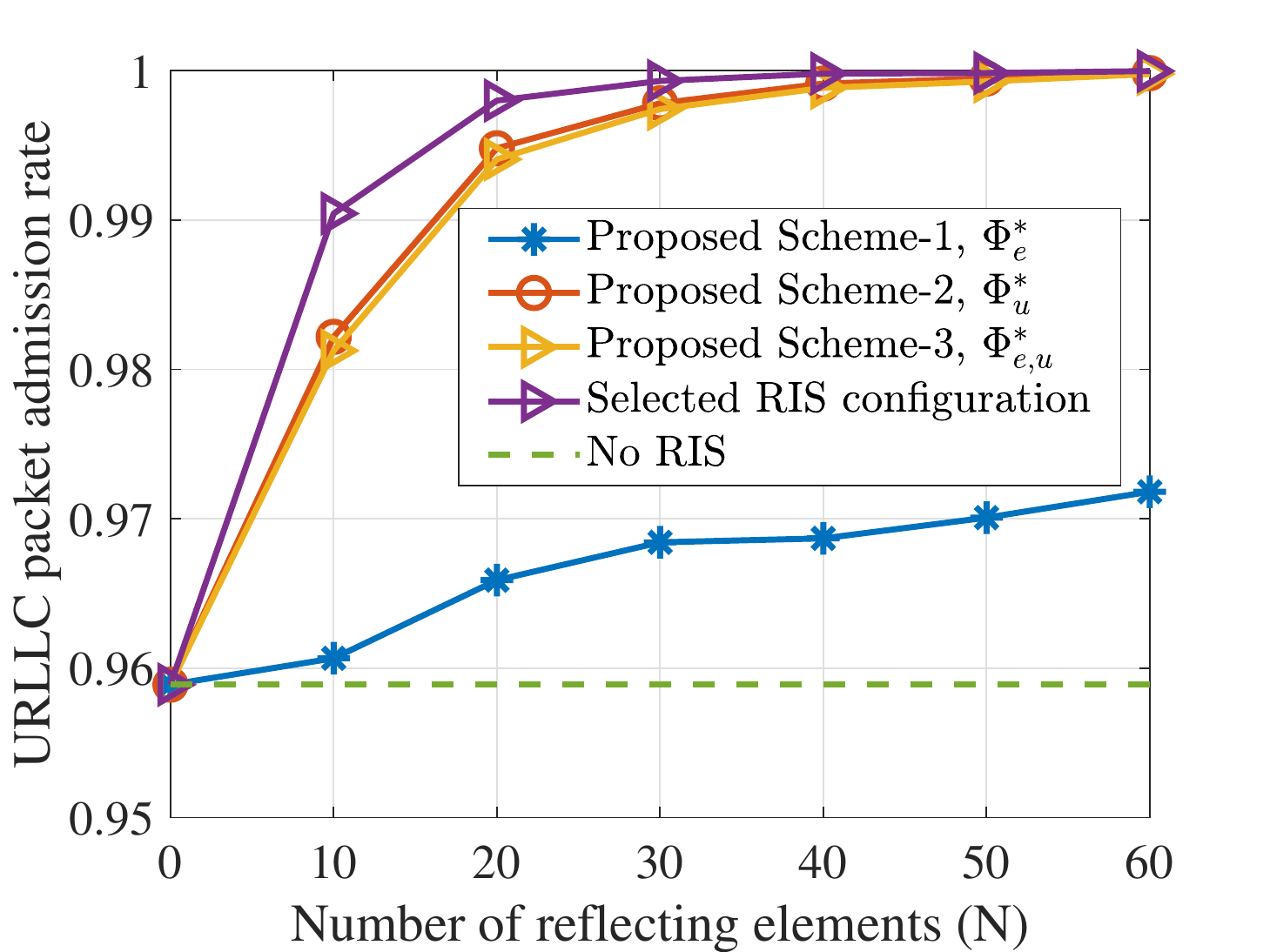}
\caption{URLLC packet admission rate against the number of RIS elements. }
\label{Fig:N_reliabiy}
\end{subfigure}~\hspace{0 in}
\begin{subfigure}[b]{.35\columnwidth}
\centering
\includegraphics[width=1\columnwidth,draft=false]{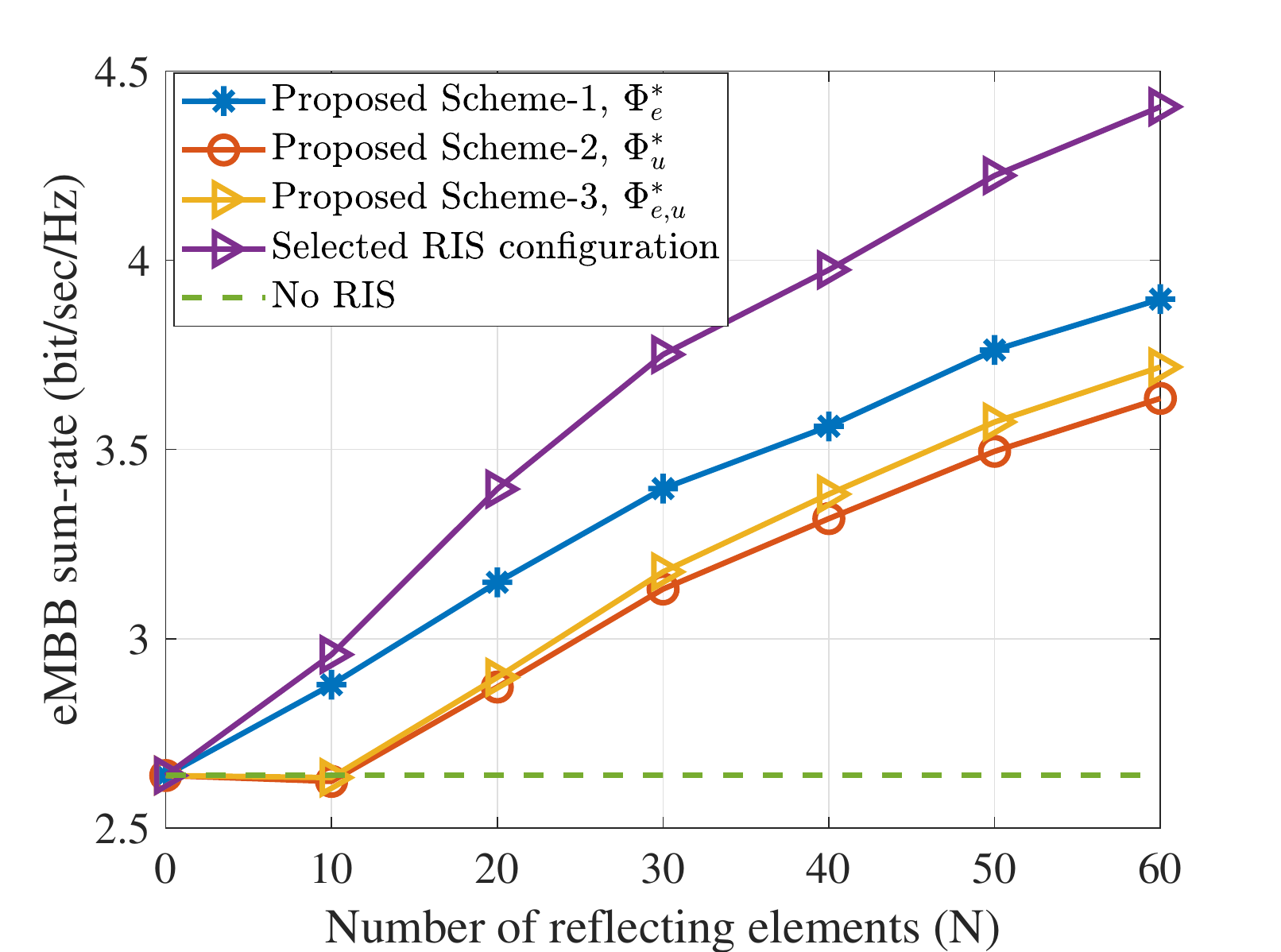}
\caption{eMBB sum rate against the number of RIS elements.}
\label{Fig:N_RATE}
\end{subfigure}
\caption{Performance of the proposed algorithm against the number of RIS elements. $\delta=0.1$ and $U=65$ }
\label{fig:N}
\vspace{-.1 in}
\end{figure}
 Fig. \ref{fig:N} depicts the impact of the size of the RIS ($N$) and the main observations are summarized below.
 \begin{itemize}
     \item It can be shown that the proposed schemes outperform the baseline with no RIS in both the URLLC packets admission rate and the eMBB sum rate. Particularly, as shown in Fig. \ref{Fig:N_reliabiy}, the proposed schemes achieve $96.87\%$, $99.9\%$, $99.9$, and $99.98\%$ URLLC packets admission rate at $N=40$ compared to $95.6\%$ when the RIS is not deployed. For the same number of reflecting elements $N=40$, Fig. \ref{Fig:N_RATE} shows that the proposed schemes achieve enhancement on the eMBB rate around $52\%$, $37\%$, $30\%$ and $27\%$ when the RIS is not deployed. The reason behind that is the ability of the BS to select a phase-shift matrix, depending on the number of the URLLC packets and their channel conditions, from the set $\{\boldsymbol{\Phi}_{\rm e}^*,\boldsymbol{\Phi}_{\rm u}^*,\boldsymbol{\Phi}_{\rm e,u}^*\}$ to configure the RIS. Particularly, the selected phase-shift matrix should achieve the best performance in terms of the URLLC packet admission rate and the minimum eMBB loss.\footnote{\textcolor{black}{It is important to note here that the achieved gain of the proposed scheme in terms of the URLLC reliability and the eMBB sum-rate comes at the expense of an extra overhead to transmit the different RIS configurations and to switch between them.}}
     \item We can also see that only $N=60$ RIS elements are enough to achieve $99.99\%$ URLLC packets admission rate along with around $70\%$ enhancement of the eMBB sum rate compared to the case when no RIS is deployed. 
     \item  The trade-off between the URLLC packet admission rate and the eMBB sum rate is clear in the behavior of scheme-1, scheme-2 and scheme-3. By enhancing the channels condition of the URLLC traffic, scheme-2 and scheme-3 give the URLLC traffic a higher priority over the eMBB traffic which means better URLLC packet admission rate and more eMBB rate-loss. Conversely, scheme-1 gives the eMBB traffic a higher priority over the URLLC counterpart by using the eMBB phase-shift matrix that was optimized to enhance the eMBB rate. This leads to a better eMBB sum rate and a lower URLLC admission.
     \item Increasing the number of RIS elements enhances both the URLLC packet admission rate and the eMBB sum rate. This is evident since at higher $N$, the URLLC channel conditions are enhanced and a higher eMBB rate is achieved, which means a higher availability for eMBB resources to allocate the URLLC packets. While increasing $N$, scheme-1 exhibits lower enhancement in URLLC service admission. In other words, unlike $\boldsymbol{\Phi}_{\rm u}^*$ and $\boldsymbol{\Phi}_{\rm e,u}^*$, $\boldsymbol{\Phi}_{\rm e}^*$ acts as a random phase-shift matrix for the URLLC traffic, i.e., the improvement on the URLLC channel conditions is moderate \cite{ghanem2020joint}. 
 \end{itemize}
 \vspace{-0.2in}
\subsection{Effect of the BS transmission power} 
 \begin{figure}[t]
\centering
\begin{subfigure}[b]{.32\columnwidth}
\centering
\includegraphics[width=1\columnwidth,draft=false]{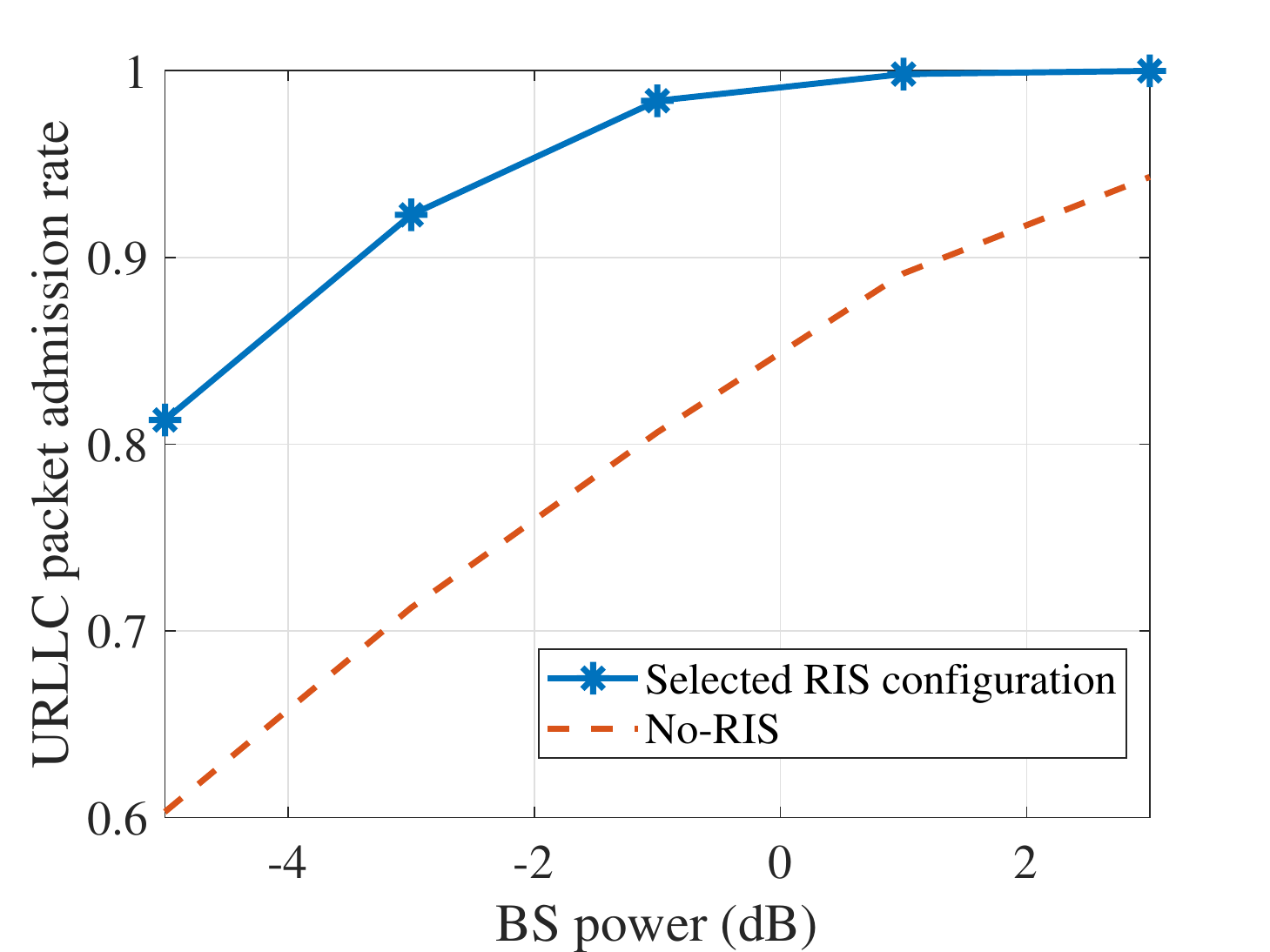}
\caption{URLLC packet admission rate against the BS transmission power. }
\label{Fig:p_reliabiy}
\end{subfigure}
~\hspace{-.1 in}
\begin{subfigure}[b]{.32\columnwidth}
\centering
\includegraphics[width=1\columnwidth,draft=false]{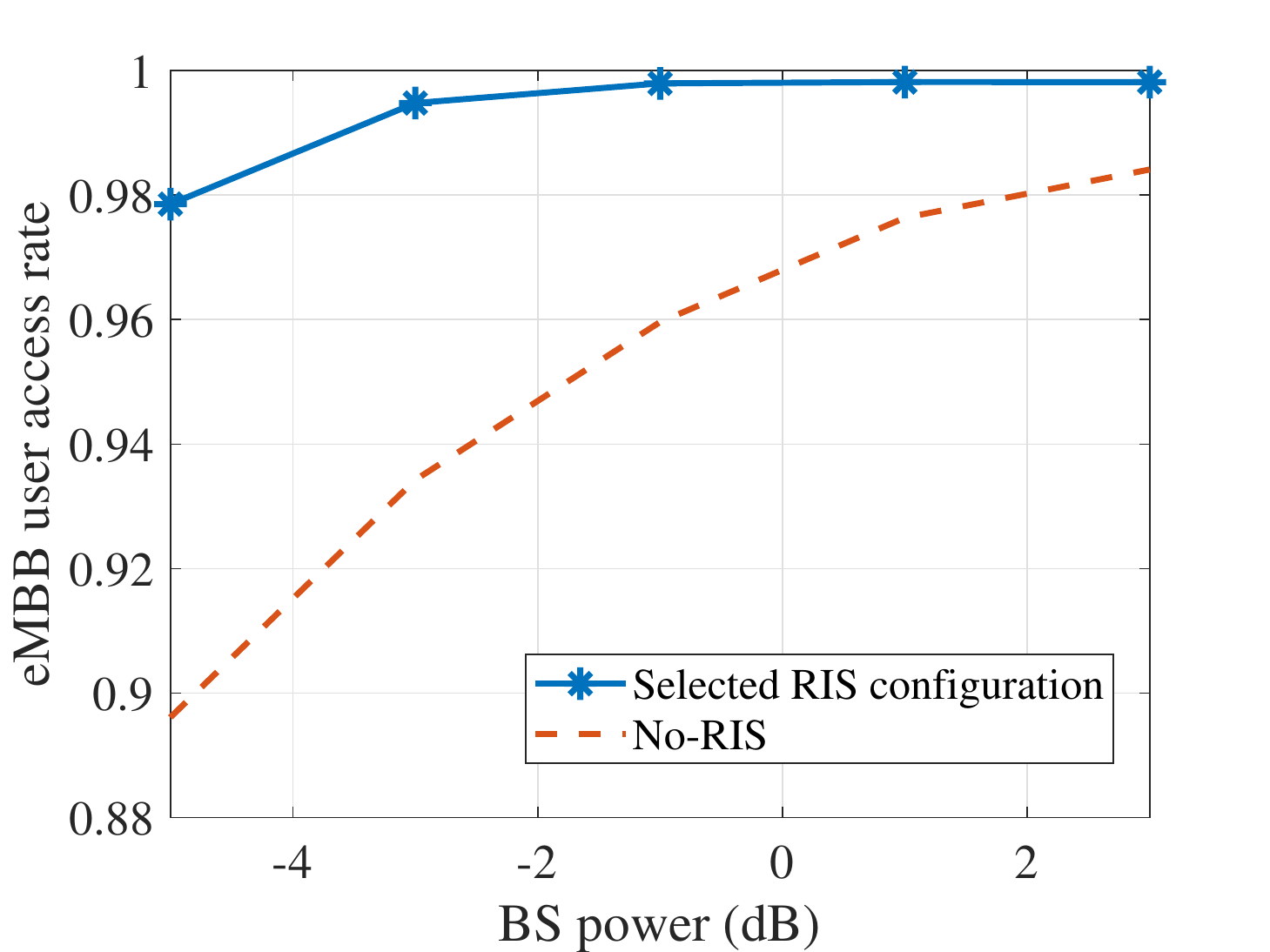}
\caption{eMBB users admission rate against the BS transmission power.}
\label{Fig:p_acc}
\end{subfigure}
~\hspace{-.1 in}
\begin{subfigure}[b]{.32\columnwidth}
\centering
\includegraphics[width=1\columnwidth,draft=false]{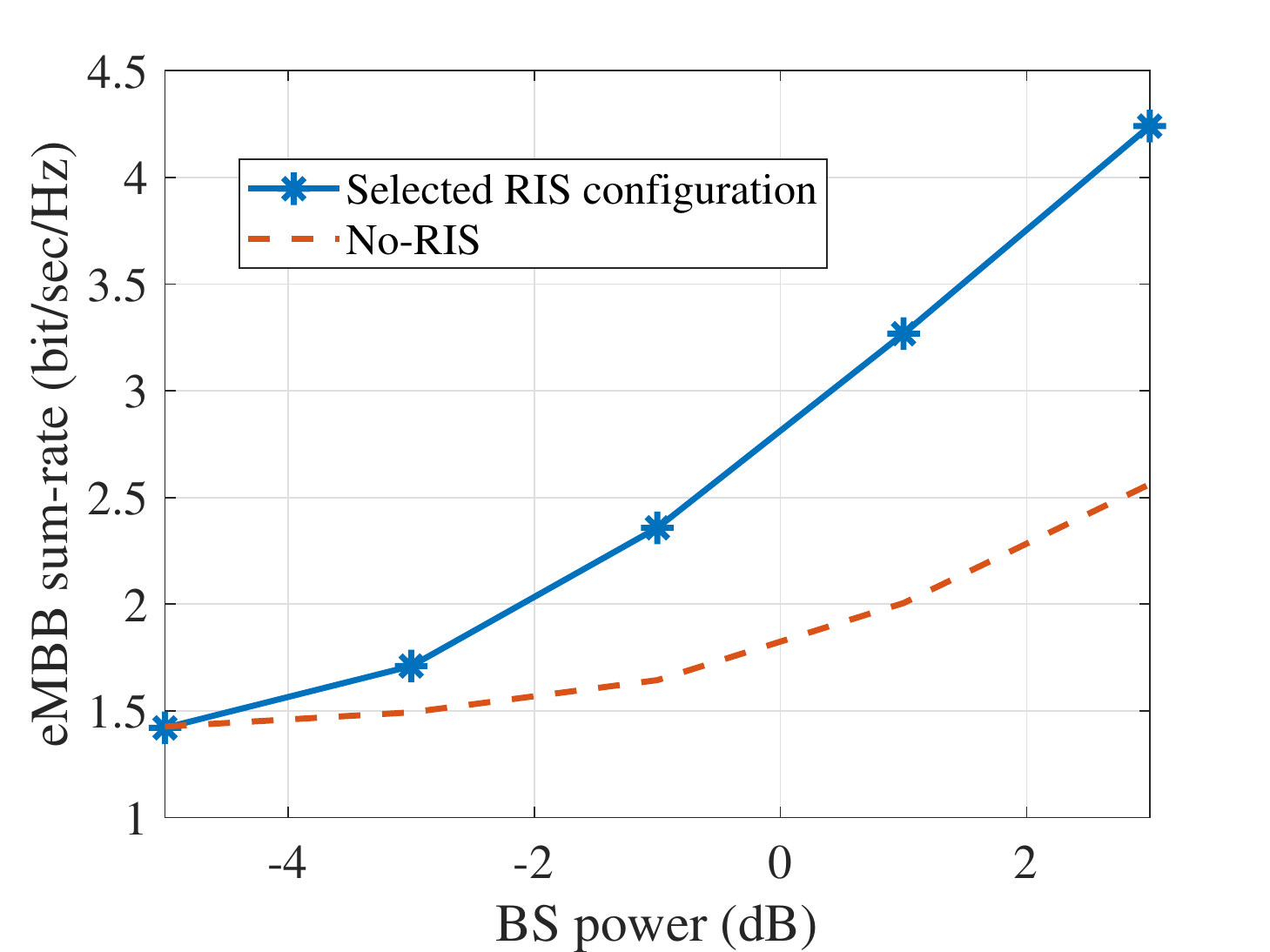}
\caption{eMBB sum rate against the BS transmission power.}
\label{Fig:p_RATE}
\end{subfigure}
\caption{Performance of the proposed algorithm against the BS transmission power. $N=50$ and $U=65$. }
\label{fig:p}
\end{figure}
Fig. \ref{fig:p} depicts the impact of of the BS transmission power $P_{\rm BS}$; the observations on Fig. \ref{fig:p} are summarized as follows. 
 \begin{itemize}
     \item As shown in Fig. \ref{Fig:p_reliabiy} and Fig. \ref{Fig:p_acc}, the selected RIS configuration can achieve better URLLC packet admission rate and better eMBB users admission rate compared to the case of no RIS is deployed. As whown in Fig. \ref{Fig:p_reliabiy}, the achieved gain of using the RIS is equivalent to $4$ dB on the transmission power at $90\%$ URLLC packet admission rate. This is attributed to the benefits of RIS in enhancing the channel conditions of the URLLC users such that the URLLC load can be admitted with limited frequency and power resources. On the other hand, the gain is equivalent to $4.5$ dB at eMBB users admission rate of $95$\%. Moreover, by increasing the transmission power, the URLLC packet admission rate and the eMBB users admission rate enhance because more power resources are available to guarantee the URLLC and the eMBB rates requirements. By keeping increasing the transmission power, the gain achieved by the proposed scheme is reduced on the URLLC packet admission rate and the eMBB users admission rate. However this decreasing is translated to better gain on the eMBB sum rate as shown in Fig. \ref{Fig:p_RATE}.
     \item Fig. \ref{Fig:p_RATE} shows that at low transmission power the selected RIS configuration has a quite similar performance to the baseline in terms of the eMBB sum rate, whereas a better behaviour can be seen at high transmission power. This is attributed to the benefits of RIS in enhancing the URLLC admission which means less eMBB resources are punctured to allocate higher URLLC load.  
 \end{itemize}
\subsection{Effect of the rate margin factor $\delta$} 
\begin{figure}[t]
\centering
\begin{subfigure}[b]{.32\columnwidth}
\centering
\includegraphics[width=1\columnwidth,draft=false]{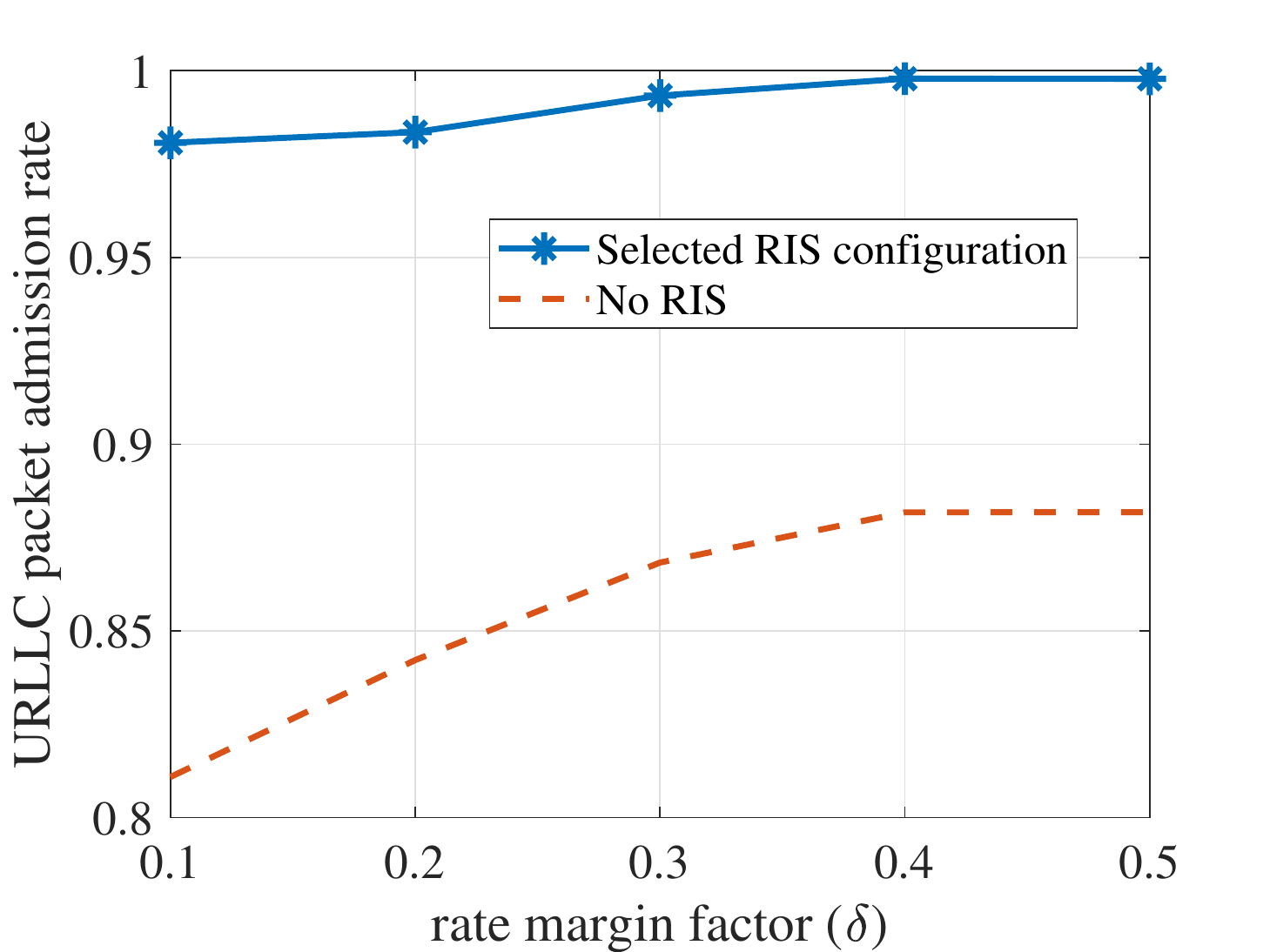}
\caption{URLLC packet admission rate against the rate margin factor. }
\label{Fig:dis_reliabiy}
\end{subfigure}
~\hspace{-.1 in}
\begin{subfigure}[b]{.32\columnwidth}
\centering
\includegraphics[width=1\columnwidth,draft=false]{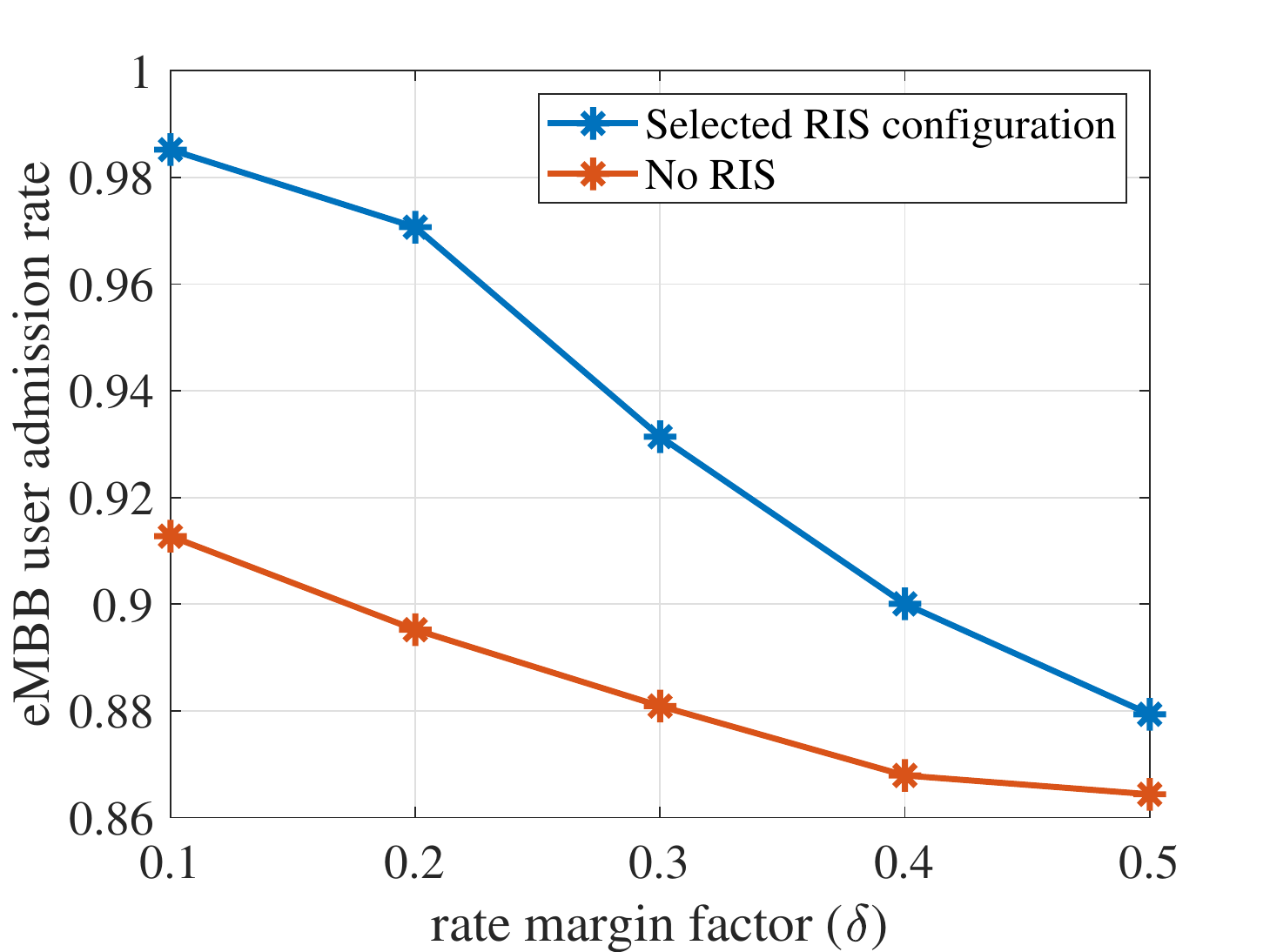}
\caption{eMBB users admission against the rate margin factor.}
\label{Fig:delta_acc}
\end{subfigure}
~\hspace{-.1 in}
\begin{subfigure}[b]{.32\columnwidth}
\centering
\includegraphics[width=1\columnwidth,draft=false]{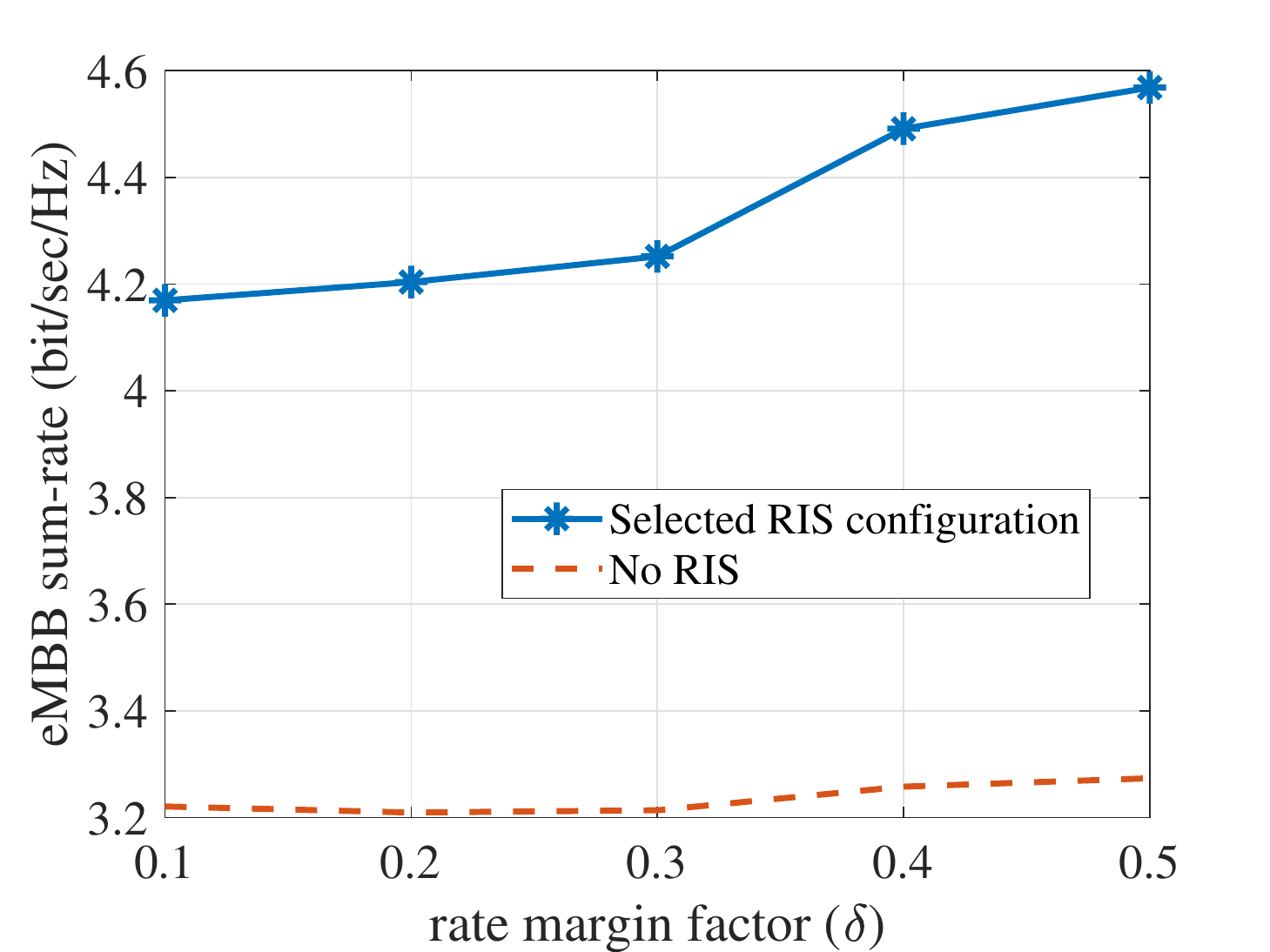}
\caption{eMBB sum rate against the rate margin factor.}
\label{Fig:dist_RATE}
\end{subfigure}
\caption{Performance of the proposed algorithm against rate margin. $N=50$, $U=65$ and $r_{\rm th}=7$ Mbps. }
\label{fig:delta}
\end{figure}
\begin{figure}[tb!]
\centering
\begin{subfigure}[b]{.45\columnwidth}
\centering
\includegraphics[width=0.8\columnwidth,draft=false]{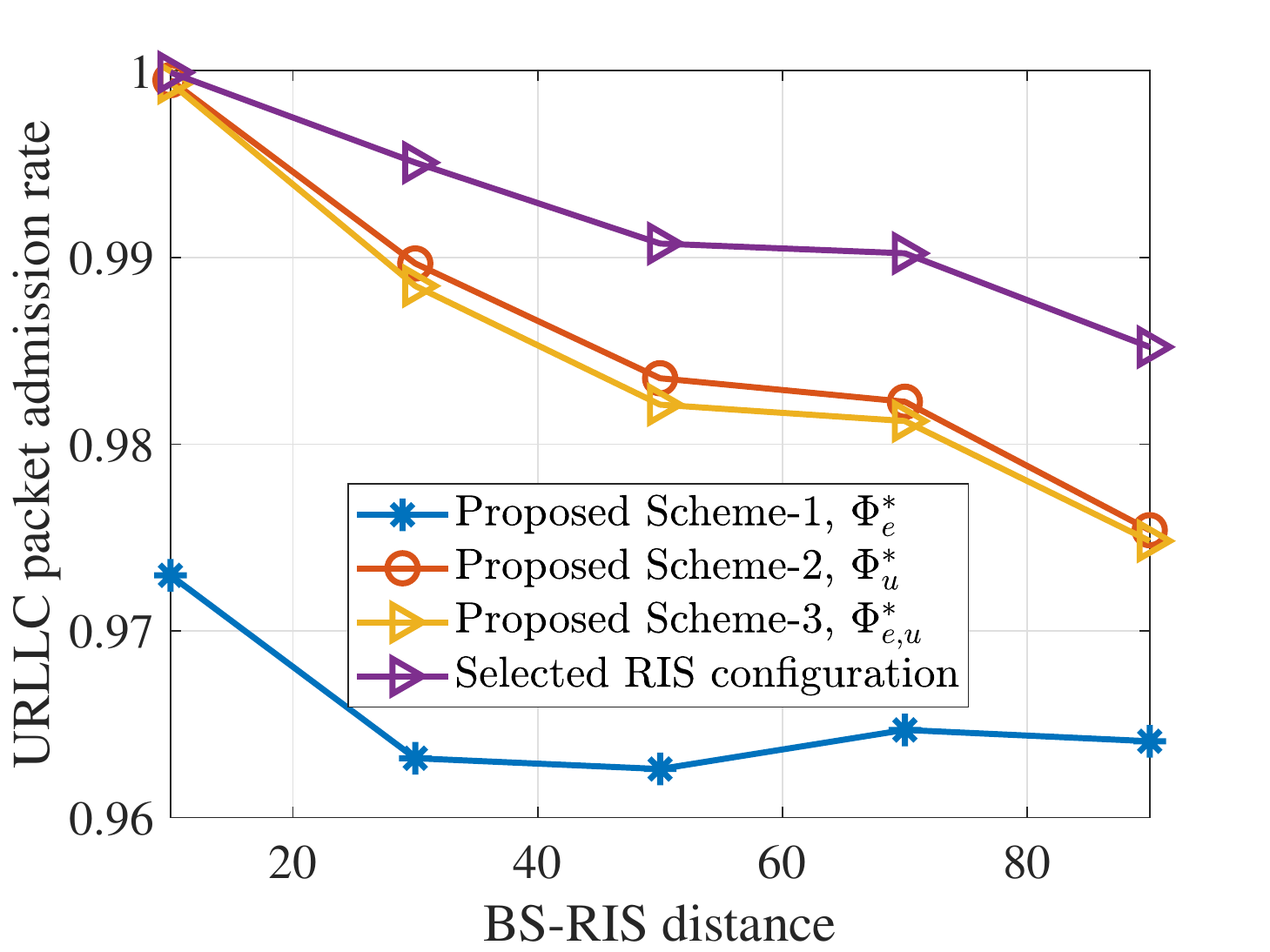}
\caption{URLLC packets admission rate versus the distance between the BS and the RIS. }
\label{Fig:delta_reliabiy}
\end{subfigure}
~\hspace{-.1 in}
\begin{subfigure}[b]{.45\columnwidth}
\centering
\includegraphics[width=0.8\columnwidth,draft=false]{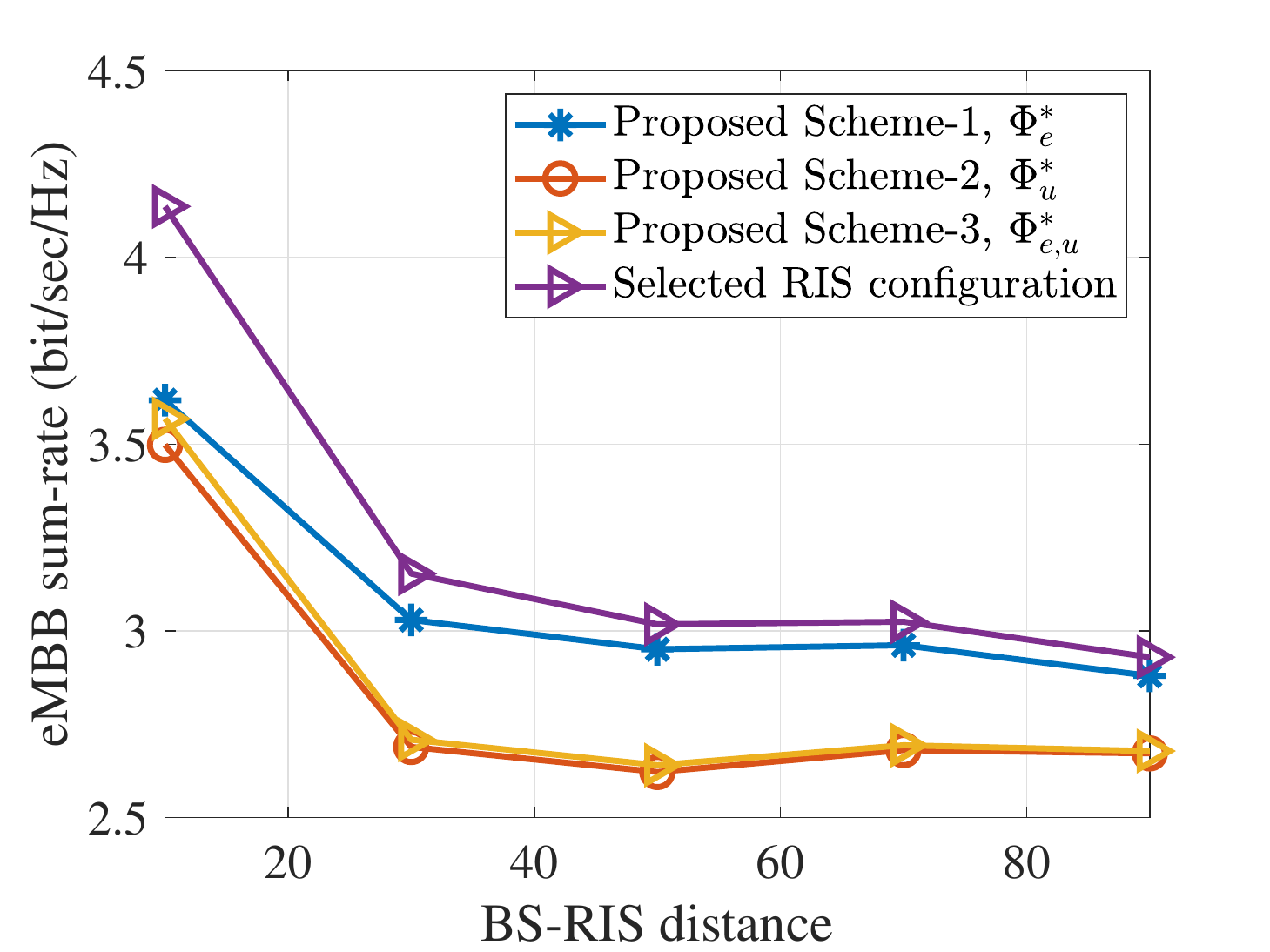}
\caption{eMBB sum rate versus the distance between the BS and the RIS. \textcolor{white}{eMBB sum rate}}
\label{Fig:delta_RATE}
\end{subfigure}
\caption{Performance of the proposed algorithm versus the distance between the BS and the RIS, where $N=20$ and $U=65$. }
\label{fig:dist}
\end{figure}

Fig. \ref{fig:delta} illustrates the impact of the rate margin factor $\delta$ on the URLLC packets admission rate, the eMBB users admission rate, and the eMBB sum rate. This figure shows that increasing the margin factor enhances the URLLC packets admission and reduces the eMBB users admission. This is because increasing $\delta$ means more over-provisioned resources for the URLLC traffic, i.e., higher rate per eMBB user than the requested, which makes them later available to allocate the URLLC load. Hence, the enhanced URLLC packet admission rate. However, when higher rates are attained per eMBB user, this implies lower eMBB admission rate allowing only eMBB users with good channel conditions to be admitted, which will increase the sum rate, as shown in Fig. \ref{fig:delta}.c. Further, as shown in Fig. \ref{fig:delta}.b, increasing $\delta$ causes drastic reduction in the eMBB admission, since the network will over-provision much of its resources to satisfy the rate constraint per user and the RIS will be optimized to serve only such users with good channel conditions, leaving many of the eMBB users not admitted. This however helpsthe  URLLC service to attain a better QoS.
\textcolor{black}{
\subsection{Effect of the RIS Location} 
Fig. \ref{fig:dist} illustrates the impact of varying the distance between the BS and the RIS on the URLLC packet admission rate and the eMBB sum rate. This figure shows that locating the RIS away from the BS impacts the performance of both URLLC and eMBB services in terms of the URLLC admission and the sum rate of eMBB, respectively. Indeed, this is consistent with other findings in the literature about the RIS deployment being most beneficial when it is located near the BS or close to the user \cite{8888223}. However, the users are located randomly in the network. Hence, the best location of the RIS in this context is to be closer to the BS, which represents the optimal RIS placement for all users coexisting in the network.}
\section{Conclusions}\label{Conclusion}
In this paper, we studied the RIS technology for enabling the coexistence of eMBB and URLLC services in a wireless networks. The RIS is deployed to improve the performance of the URLLC and eMBB users by controlling their channel conditions. Two optimization problems are formulated for multiplexing the eMBB and URLLC traffic, i.e., the time-slot basis eMBB allocation problem and the mini-time slot URLLC allocation problem. The eMBB allocation problem has the objective of maximizing the eMBB sum rate while satisfying the eMBB QoS requirements. Meanwhile, the URLLC allocation problem aims at maximizing the admitted URLLC packets and minimizing the eMBB rate loss while satisfying the QoS of eMBB and URLLC. To overcome the high computational complexity of optimizing the RIS phase-shift matrix per mini-time slot, we proposed a proactively designed RIS phase-shift matrices that are optimized at the beginning of the time slot. Simulation results show that the proposed algorithm has low time complexity which makes it a practical scheme to delay-sensitive URLLC traffic. It is also shown that the proposed RIS scheme achieves 99.99\% URLLC packet admission rate using only $60$ RIS elements. Moreover, the proposed model can achieve up to 70\% enhancement on the eMBB rate compared to no-RIS is deployed. \textcolor{black}{The proposed approach presents a general framework of RIS-aided eMBB-URLLC traffic multiplexing. Nevertheless, the proposed scheme can be extended for MISO/MIMO, NOMA, and multi-cell scenarios.  Furthermore, machine learning techniques can be integrated in these setups to reduce the complexity of the RIS phase-shift matrix optimization. These problems can be considered as potential future research directions.}
\appendices 
\section{Solution Approach of Problem $\mathcal{P}_{1,2}$}
Let us start by defining $\boldsymbol{\vartheta}^H=[\vartheta_1, \vartheta_2, \dots, \vartheta_N]^H$, where $\vartheta_n = e^{j\phi_n}$. Therefore, we obtain $ |h_{{\rm BS}, e} +  \mathbf{{h}}_{{\rm RIS},e}^H\boldsymbol{\Phi} \,\mathbf{f}_{{\rm  BS},{\rm  RIS}}|^2 = |h_{{\rm BS}, e} + \boldsymbol{\vartheta}^H\,\boldsymbol{\Theta}|^2 = \boldsymbol{\vartheta}^H\,\boldsymbol{\Theta}\,\boldsymbol{\Theta}^H\boldsymbol{\vartheta}  + h_{{\rm BS}, e}\,\boldsymbol{\Theta}^H\,\boldsymbol{\vartheta}+\boldsymbol{\vartheta}^H\,\boldsymbol{\Theta}\,h_{{\rm BS}, e}^\dagger\,+|h_{{\rm BS}, e}|^2$, where $\boldsymbol{\Theta} = {\rm diag}(\mathbf{{h}}_{{\rm RIS},e}^H)\,\mathbf{f}_{{\rm  BS},{\rm  RIS}} $, and $\boldsymbol{\Theta} \in \mathbb{C}^{N \times 1}$. By introducing an auxiliary variable $\rho$, problem $\mathcal{P}_{1,2}$ can be equivalently transformed to 
\begin{subequations}\label{phaseshiftoptimization_2}
\begin{align}
  {\mathcal{P}_{1,3}}: \quad &\max_{\Bar{\boldsymbol{\vartheta}}}\,\, \sum_{e=1}^E \log_2\left(1+\frac{p_e\,(\Bar{\boldsymbol{\vartheta}}^H\,\boldsymbol{Q}_e\,\Bar{\boldsymbol{\vartheta}} + |h_{{\rm BS}, e}|^2)}{\Gamma_{\rm eMBB}\,\sigma^2}\right)\label{PS2_obj}\\
  \text{s.t.} \quad & \left(1-\delta\right)\log_2\left(1+\frac{p_e\,(\Bar{\boldsymbol{\vartheta}}^H\,\boldsymbol{Q}_e\,\Bar{\boldsymbol{\vartheta}}+|h_{{\rm BS}, e}|^2)}{\Gamma_{\rm eMBB}\,\sigma^2}\right)\ge \frac{r_{\rm th}}{W\,b}, \forall\,\, \qquad e \in \mathcal{E}_f,\label{PS2_C1}\\
   & \qquad |\Bar{\boldsymbol\vartheta}_n|=1,\qquad \forall n = 1,\dots N+1  \label{PS2_C3},
\end{align}
\end{subequations}
where 
\begin{equation}\label{PSymtric_matrix}
    \boldsymbol{Q}_e =
\begin{bmatrix}
\boldsymbol{\Theta}\,\boldsymbol{\Theta}^H &  \boldsymbol{\Theta}\,h_{{\rm BS}, e}^\dagger\\
h_{{\rm BS}, e}\,\boldsymbol{\Theta}^H& 0 
\end{bmatrix}, \quad \text{and} \quad
\Bar{\boldsymbol{\vartheta}}=
\begin{bmatrix}
\boldsymbol\vartheta\\
\rho
\end{bmatrix},
\end{equation}
such that $\Bar{\boldsymbol{\vartheta}}^H\,\boldsymbol{Q}_e\,\Bar{\boldsymbol{\vartheta}}={\rm tr}(\boldsymbol{Q}_e\,\Bar{\boldsymbol{\vartheta}}\,\Bar{\boldsymbol{\vartheta}}^H)$. In addition, we define $\boldsymbol{S}=\Bar{\boldsymbol{\vartheta}}\Bar{\boldsymbol{\vartheta}}^H$, which needs to satisfy $\mathrm{rank}(\Bar{\boldsymbol{\vartheta}}) = 1$. This rank one constraint is a non-convex constraint \cite{elhattab2021reconfigurable}. By dropping this constraint, we reach 
\begin{subequations}
\label{phaseshiftoptimization_3}
\begin{align}
  \mathcal{P}_{1,4}: \quad &\max_{\boldsymbol{S}}\,\, \sum_{e=1}^{E_f} \log_2\left(1+\frac{p_e\,({\rm tr}(\boldsymbol{Q}_e\boldsymbol{S})+|h_{{\rm BS}, e}|^2)}{\Gamma_{\rm eMBB}\,\sigma^2}\right)\label{PS3_obj}\\
  \text{s.t.} \quad & \left(1-\delta\right)\,b\,\log_2\left(1+\frac{p_e\,({\rm tr}(\boldsymbol{Q}_e\boldsymbol{S})+|h_{{\rm BS}, e}|^2)}{\Gamma_{\rm eMBB}\,\sigma^2}\right)\ge {r_{\rm th}}, \quad \forall\,\, e \in \mathcal{E}_f,\label{PS3_C1}\\
   &\boldsymbol{S}_{n,n}=1,\qquad \qquad \qquad \qquad  \qquad \qquad \qquad \qquad \qquad \,\,\,\,  \forall n \in \left\{1,2,\dots,N+1\right\}  \label{PS3_C3},\\
   &\boldsymbol{S}\succeq 0.
\end{align}
\end{subequations}
\indent It can be easily seen that problem $\mathcal{P}_{1,4}$ is a semi-definite programming (SDP) problem, which can be optimally solved using one of the convex optimization solvers such as CVX \cite{cvx}. In general, the optimal $\Bar{\boldsymbol\vartheta}$ obtained by solving problem $\mathcal{P}_{1,4}$ does not satisfy the rank-one constraint \cite{elhattab2021reconfigurable}. Consequently, the Gaussian randomization technique is applied to get a rank-one solution \cite{elhattab2021reconfigurable}. 
\section{Solution Approach of Problems $\mathcal{P}_{3}$ and $\mathcal{P}_{4}$} 
\label{RIS-URLLC}
We provide here the solution for problems $\mathcal{P}_3$ and $\mathcal{P}_4$. Problems $\mathcal{P}_3$ and $\mathcal{P}_4$ (which aim at maximizing the URLLC channels and URLLC-eMBB channels respectively)  have similar formulation. We start by solving $\mathcal{P}_3$. By adding a auxiliary variable $\zeta$, $\mathcal{P}_3$ is re-written as shown in problem $\mathcal{P}_{3,1}$ in \eqref{P31} on top of next page. Then, using the same results in \eqref{phaseshiftoptimization_3}, problem $\mathcal{P}_{3,1}$ is re-written as shown in problem $\mathcal{P}_{3,2}$ in \eqref{P32} on top of next page. Finally, problem $\mathcal{P}_4$ can be solved by following the same steps as in problem $\mathcal{P}_3$.
\begin{figure*}
\begin{minipage}{.5\textwidth}
\begin{subequations}
\label{P31}
\begin{align}
  \mathcal{P}_{3,1}:\quad  &\max_{\zeta,\boldsymbol{\Phi}_{u}}\zeta\\
 \text{s.t.}\quad  &|g_{{\rm BS}, u}+  \boldsymbol{{g}}_{{\rm RIS},u}^H \boldsymbol{\Phi}_{u} \,\mathbf{f}_{{\rm  BS},{\rm  RIS}}|^2\ge\zeta\,\, \forall u\in \mathcal{U},\\
 &0\le\phi_n\le 2\pi,\quad \,\,\forall n=1,\dots,N+1\\
 &\zeta\ge 0.
\end{align}
\end{subequations}
\end{minipage}
\begin{minipage}{.5\textwidth}
\begin{subequations}\label{P32}
\begin{align}
  {\mathcal{P}}_{3,2}: \quad &
 \max_{\zeta}\zeta \\
  \text{s.t.} \quad &{\rm tr}(V_uS)+|g_{{\rm BS}, u}|^2\ge \zeta,\,\, \forall u\in \mathcal{U},\\
    &  S_{n,n}=1,\,\,\forall n=1,\dots,N+1,\\
   &S\succeq 0,\\
   &\zeta\ge 0.
\end{align}
\end{subequations}
\end{minipage}
\noindent\makebox[\textwidth]{\rule{\textwidth}{0.4pt}}
\end{figure*}
\bibliographystyle{IEEEtran}
\bibliography{main.bib}

\end{document}